\documentclass[runningheads]{llncs}
\usepackage[T1]{fontenc}
\usepackage{graphicx}
\usepackage{amsfonts}
\usepackage{stmaryrd}
\usepackage{etoolbox}
\usepackage{mathpartir}
\usepackage{mathtools}
\usepackage{xcolor}
\usepackage[noend]{algpseudocode}
\usepackage{algorithm}
\usepackage{sty/prooftree}
\usepackage{hyperref}
\usepackage{amsmath} 
\usepackage[capitalize]{cleveref}
\usepackage{centernot}
\usepackage[all]{xy} 

\newcommand{\refItem}[2]{\Cref{#1}~(\ref{#1:#2})}

\newcommand{\fun}[3]{\ensuremath{#1 : #2 \rightarrow #3}} 
\newcommand{\N}{\mathbb{N}} 
\newcommand{\ple}[1]{\ensuremath{\langle #1 \rangle}} 
\newcommand{\wpf}{\wp_F} 

\newcommand{\sem}[1]{\llbracket #1 \rrbracket}

\def\eg{\emph{e.g. }}
\def\ie{\emph{i.e. }}

\spnewtheorem{app-thm}{Theorem}[section]{\bfseries}{\itshape}

\crefname{figure}{Figure}{Figures}
\crefname{theorem}{Theorem}{Theorems}
\crefname{lemma}{Lemma}{Lemmas}
\crefname{property}{Property}{Properties}
\crefname{corollary}{Corollary}{Corollaries}
\crefname{app-thm}{Theorem}{Theorems}

\newcommand{\Events}{\mathcal{E}} 
 
\newcommand{\PP}{P}
\newcommand{\PQ}{Q} 
\newcommand{\Vtre}{\mathcal{V}_3} 
\newcommand{\Vsei}{\mathcal{V}_6} 
\newcommand{\mnt}{M} 
\newcommand{\gmnt}{\widehat{\mnt}}  
\newcommand{\amnt}[1]{\mathcal{M}_{#1}} 
\newcommand{\gamnt}[1]{\widehat{\mathcal{M}}_{#1}} 
\newcommand{\yv}{\mathsf{yes}}
\newcommand{\nv}{\mathsf{no}}
\newcommand{\uv}{\mathsf{?}} 
\newcommand{\uyv}{\uv_\yv}
\newcommand{\unv}{\uv_\nv}
\newcommand{\nmv}{\mathsf{\chi}} 
\newcommand{\SF}{\mathbb{S}}
\newcommand{\coSF}{{\mathsf{co}\SF}}

\newcommand{\clop}[1][]{\Gamma_{#1}} 
\newcommand{\inop}[1][]{\Delta_{#1}} 
\newcommand{\NR}{\mathsf{NR}}

\newcommand{\tr}{\sigma}

\newcommand{\ftr}{u}
\newcommand{\aftr}{v} 
\newcommand{\prefix}{\triangleleft} 

\newcommand{\AP}{AP}
\newcommand{\pp}{p}

\newcommand{\pq}{q}
\newcommand{\ev}{e}
\newcommand{\psem}[1]{\langle\!\langle #1 \rangle\!\rangle} 
\newcommand{\Ftr}{\Events^\star}
\newcommand{\Itr}{\Events^\omega}
 
\newcommand{\itr}{w}

\newcommand{\ff}{\phi}
\newcommand{\fp}{\psi}
\newcommand{\until}{\mathrel{\mathcal{U}}}
\newcommand{\release}{\mathrel{\mathcal{R}}}

\newcommand{\F}{\mathop\Diamond}
\newcommand{\G}{\mathop\square}
\newcommand{\Next}{\mathop{\circ}}
\newcommand{\mods}{\mathrel{\models}}
\newcommand{\lmods}{\mods_L}
\newcommand{\lsem}[1]{\sem{#1}}

\renewcommand{\tt}{\mathsf{t}}
\renewcommand{\ts}{\mathsf{s}}
\newcommand{\co}[1]{#1^\bot}
\newcommand{\rec}[2]{\nu #1.#2}
\newcommand{\dual}[1]{\overline{#1}}
\newcommand{\subst}[2]{\{#1/#2\}}
\newcommand{\cmods}{\mods_C}
\newcommand{\csem}[1]{\sem{#1}}

\newcommand{\red}[1]{\stackrel{#1}\rightarrow}
\newcommand{\reds}[1]{\stackrel{#1}\Rightarrow_{*}}
\newcommand{\ireds}[1]{\stackrel{#1}\Rightarrow_{\omega}}

\newcommand{\var}{\mathsf{X}}
\newcommand{\vary}{\mathsf{Y}}

\newcommand{\rank}[1]{\mathcal{R}(#1)}
\newcommand{\fn}[1]{fn(#1)}

\newcommand{\prf}[1]{\vdash #1}
\newcommand{\Ctx}{\Gamma}
\newcommand{\CtxD}{\Delta}

\newcommand{\drv}[3]{#1 \vdash #2 \mapsto #3}

\newcommand{\Buchi}{Büchi }
\newcommand{\Automaton}{\mathcal{A}}
\newcommand{\NBA}{\Automaton}
\newcommand{\NFA}{\hat\Automaton}
\newcommand{\DFA}{\tilde\Automaton}
\newcommand{\FairA}[1]{\lceil #1 \rceil}
\newcommand{\Language}[1]{\mathcal{L}(#1)}
\newcommand{\tuple}[1]{<#1>}
\newcommand{\States}{\mathcal{Q}}
\newcommand{\Alphabet}{\Sigma}
\newcommand{\act}{\alpha}
\newcommand{\actb}{\beta}

\newcommand{\trans}{\mathcal\delta}
\newcommand{\IStates}{\States_0}
\newcommand{\FStates}{\mathcal{F}}

\newcommand{\PFstar}[1]{\wpf^*(#1)}
\newcommand{\st}{q}
\newcommand{\move}[1]{\red{#1}}
\newcommand{\moves}[1]{\reds{#1}}
\newcommand{\imoves}[1]{\ireds{#1}}
\newcommand{\SSet}{S}
\newcommand{\Tact}[1]{T(#1)}
\newcommand{\Talg}[1]{T(#1)}
\newcommand{\Tqalg}[2]{T(#1,#2)}

\newcommand{\spec}{S}

\newcommand{\rulename}[1]{\textup{\textsc{\small[#1]}}}
\newcommand{\defrule}[1]{\hypertarget{rule:#1}{\rulename{#1}}}
\newcommand{\refrule}[1]{\hyperlink{rule:#1}{\rulename{#1}}}
\newcommand{\proofcase}[1]{\textit{#1}.}
\newcommand{\proofrule}[1]{\proofcase{Case \refrule{#1}}}
\newcommand{\set}[1]{\{#1\}}

\newcommand{\eqdef}{\stackrel{\text{\tiny\sf def}}=}

\newcommand{\Tree}[1]{\mathsf{Tr}^\infty(#1)}
\newcommand{\tree}{t}

\usepackage{sty/monitorability} 

\begin{document}
%
\title{Ain't No Stoppin' Us Monitoring Now}
%
%
%
\author{
Luca Ciccone\inst{1}\orcidID{0000-0001-9515-5280}
\and 
Francesco Dagnino\inst{2}\orcidID{0000-0003-3599-3535}
\and
Angelo Ferrando\inst{2}\orcidID{0000-0002-8711-4670}
}
\authorrunning{L. Ciccone, F. Dagnino and A. Ferrando} 
%
\institute{
University of Turin, Italy
\and
University of Genoa, Italy
}
\maketitle              
\begin{abstract}
Not all properties are monitorable. This is a well-known fact, and it means there exist properties that cannot be fully verified at runtime. However, given a non-monitorable property, a monitor can still be synthesised, but it could end up in a state where no verdict will ever be concluded on the satisfaction (resp., violation) of the property. For this reason, non-monitorable properties are usually discarded. In this paper, we carry out an in-depth analysis on monitorability, and how non-monitorable properties can still be partially verified. We present our theoretical results at a semantic level, without focusing on a specific formalism. Then, we show how our theory can be applied to achieve partial runtime verification of Linear Temporal Logic (LTL).
\end{abstract}
%


\section{Introduction}
\label{sec:intro}

Runtime Verification (RV)~\cite{DBLP:series/lncs/BartocciFFR18,10.1145/2000799.2000800,DBLP:journals/jlp/LeuckerS09} is a well-established approach, whose aim is to 
achieve the formal verification of software (resp., hardware) systems at runtime. 
This means that, 
differently from other formal verification techniques, such as Model Checking \cite{clarke1997model}, 
RV focuses on the analysis of actual system executions, rather than on a model (\ie abstraction) of it.
More precisely, 
given a formal property, which can be expressed in various formalisms \cite{DBLP:conf/focs/Pnueli77,DBLP:journals/jacm/HennessyM85,DBLP:journals/scp/AnconaFFM21,DBLP:journals/rts/Koymans90,DBLP:conf/formats/MalerN04}, a \emph{monitor} is synthesised and used 
to verify the property against system executions. 
A monitor can be seen abstractly as a function that, given a \emph{finite observation} of a system execution 
(\eg a finite sequence, a.k.a. trace, of events generated by the system), 
returns a verdict stating whether it satisfies or violates the property. 
Intuitively, what the monitor does is checking whether the current observation of the system execution carries enough information to conclude the satisfaction (resp., violation) of the property under analysis, that is, 
it does not need to observe more to determine the verdict.

Unfortunately, for some properties, the monitor could end up in scenarios where, no matter what it observes, it would never be able to conclude a verdict. 
Such cases are problematic since the monitor keeps on observing the system, thus consuming resources, even though it has no hope to determine the satisfaction (resp., violation) of the property. 
This has lead to the notion of \emph{monitorability}~\cite{DBLP:journals/entcs/KimKLSV02,AcetoAFIL19,DBLP:series/lncs/BartocciFFR18,DBLP:conf/fm/PnueliZ06}. 
Roughly, a property is said to be \emph{monitorable} 
if we can synthesise a monitor which has always the possibility of eventually verifying it, that is, 
the above situation cannot happen. 
Actually, this is just one out of many different flavours of monitorability that exist in the literature~\cite{DBLP:journals/entcs/KimKLSV02,AcetoAFIL19,AcetoAFIL19popl,AcetoAFIL21}. 
Nevertheless, we focus on this one as 
it does not only require the existence of a correct monitor, but the latter to be well behaved as well, \ie it has not to run uselessly. 

Since monitors of non-monitorable properties may eventually be unable to conclude anything about the system, RV approaches usually rule such properties out, restricting themselves to monitorable ones. 
However, even though a non-monitorable property is in some cases ill-behaved, it may still be relevant to analyse, especially when there are fragments of it where a verdict can be reached. 

In this paper, we take seriously this idea, providing first steps towards \emph{partial} runtime verification. 
Specifically, we present an abstract semantic approach to partially monitor any given property, which makes 
our framework independent from specific formalisms used to describe properties and monitors. 
We build on the abstract setting for monitorability proposed in \cite{StuckiSSB19}. 
First of all, we observe that any property admits best safety and cosafety approximations, which turn out to be monitorable, like all safety and cosafety properties, under mild assumptions\footnote{In the standard linear time setting, safety and cosafety properties are known to be always monitorable~\cite{DBLP:journals/dc/AlpernS87,DBLP:conf/fm/PnueliZ06}, but, as we will observe, this is no longer true in our abstract setting which covers also branching time properties.}.
Then, we show how to combine the verdicts of the (standard) monitors of these approximations to build a generalised monitor, which is able to partially verify the initial property, even though it may be non-monitorable in general. 
Roughly, a generalised monitor is not only able to determine the satisfaction or violation of the property, but also when to safely give up  on the verification, since it will never be able to conclude anything. 
In a sense, we move also the monitorability check at runtime with the usual advantages: 
we do not discard a property only because we statically detect that in some cases it may not be verified, 
but rather we synthesise a monitor that, when such cases happen, stops running and, otherwise, keeps on verifying the property. 
We dub this approach partial RV, since the monitor may determine neither the satisfaction nor the violation of the property, but we are guaranteed that it keeps on running only if it has the possibility of concluding something on the property.

To test the effectiveness of the proposed framework, 
we instantiate it describing two approaches to achieve partial RV of properties in the linear time setting. 
In the first one, we focus on properties expressed in Linear Temporal Logic (LTL)~\cite{DBLP:conf/focs/Pnueli77}. 
In particular, we show how to extend the standard synthesis procedure of monitors for LTL formulas, to build a generalised monitor. 
%
In the second one, instead, we focus on the Linear Time $\nu$-calculus (a purely coinductive fragment of the Linear Time $\mu$-calculus~\cite{BarringerKP86,DBLP:conf/popl/Vardi88}). 
This calculus is much less expressive then other formalisms, such as LTL or the full Linear Time $\mu$-calculus, 
since, as we show, it can express only safety, thus monitorable, properties. 
We briefly discuss how one can directly use an $LT\nu$-term as a monitor, thus avoiding the synthesis procedure which is usually very computationally expensive, and then 
we show how to combine two $LT\nu$-terms to specify a generalised monitor. 
Finally, we prove that in this way we can describe generalised monitors of all LTL formulas. 
For both approaches we provide prototype implementations with some empirical evaluations of their performances. 

\paragraph{Outline} 
\Cref{sec:sem} revisits basic notions about monitorability in the abstract. 
Using this framework, 
\Cref{sec:completions} presents (co)safety completions of properties, 
showing how they can be used to build generalised monitors. 
\cref{sec:ltl,sec:monitor-nu} describe two instantiation of the general framework to concretely specify generalised monitors for linear time properties. 
Finally, 
\cref{sec:related} discusses related work and 
\Cref{sec:conclusions-future-work} concludes the paper.

\section{Preliminaries on monitorability} 
\label{sec:sem}

In this section, we recall well-known notions about monitorability, phrasing them in an abstract, purely semantic setting, following \cite{StuckiSSB19}. 
We need this abstract setting to develop our results independently from specific choices of system models and  specification languages. 

Assume a set $\BB$ of \emph{(system) behaviours} $\bha,\bhb,\bhc$. 
These can be whatever one needs, e.g., traces of events, trees, states, etc. 
Then, a \emph{property} is a subset of behaviours $\PP\subseteq\BB$. 
Intuitively, one cannot directly access and manipulate system behaviours as a whole, because they may have  an infinite nature; 
instead we can analyse the system through \emph{observations}, which are intuitively finite approximations of system behaviours. 

\begin{definition}[Observation system] \label{def:obs}
An \emph{observation structure} over $\BB$ consists of a preordered set \ple{\Obs,\obsord} of \emph{observations} together with a binary relation
$\bhord \subseteq \Obs\times\BB$ such that the following properties hold: 
\begin{itemize} 
\item $\obsa\obsord\obsb$ and $\obsb\bhord\bha$ implies $\obsa\bhord\bha$
\item for all $\obsa\in\Obs$ there is $\bha\in\BB$ such that $\obsa\bhord\bha$
\end{itemize} 
\end{definition}
Let us set $\BB(\obsa) = \set{ \bha \in \BB \mid \obsa\bhord\bha }$ and 
$\Obs(\bha) = \set{ \obsa\in\Obs \mid \obsa\bhord\bha }$. 
Intuitively, the preorder on observations models a refinement between them: $\obsa\obsord\obsb$ means that $\obsb$ is a finer observation w.r.t. $\obsa$. 
The relation $\bhord$ relates observations to behaviours: $\obsa\bhord\bha$ means that $\obsa$ approximates the behaviour $\bha$. 

\begin{example}[Linear Time] \label{ex:ltobs} 
Assume a set of events $\Events$. 
In the standard setting of runtime verification, 
we are interested in monitoring properties of single executions of a system. 
Hence, in this setting, 
we can take as set of behaviours either the set $\Events^\omega$ of infinite traces or the set $\Events^\infty$ of finite or infinite traces. 
Then, the set of observations can be  the set $\Events^\star$ of finite traces and both relations $\obsord$ and $\bhord$ are given by prefixing. 
\end{example}

\begin{example}[Hyperproperties]\label{ex:hyobs}
Assume again a set of events $\Events$. 
In some cases, we may be interested in monitoring properties on multiple executions of  a system. 
In this case, we can consider as set of behaviours  the set $\wp(\Events^\omega)$ (resp. $\wp(\Events^\infty)$) of sets of infinite (resp. finite or infinite) traces of events. 
Here, the set of observation can be $\wpf(\Events^\star)$ consisting of finite sets of finite traces of events, where the refinement order is given by 
$X\obsord Y$ iff  for all $\ftr \in X$ there is $\aftr \in Y$ such that $\ftr$ is a prefix of $\aftr$. 
Finally, the relation $\bhord$ relating observations and behaviours will be given by 
$X \bhord T$ iff for all $\ftr \in X$ there is $\tr \in T$ such that $\ftr$ is a prefix of $\tr$. 
Note that, a property in this setting is what is usually called \emph{hyperproperty}~\cite{ClarksonS08}. 
\end{example}

\begin{example}[Branching Time]\label{ex:btobs} 
Assume again a set of events $\Events$. 
In a branching time setting, we can take as set of behaviours the set $\Tree\Events$ of finite and infinite trees with edges labelled by events in $\Events$. 
We illustrate two possible observation structures.
A first choice for observations can be the set of finite traces $\Events^\star$ with the prefixing ordering. 
The approximation relation is given by $\ftr \bhord \tree$ iff 
$\tree$ contains a path starting from the root labelled by $\ftr$. 
Another choice for observations can be the set $\wpf(\Events^\star)$ with the structure of \cref{ex:hyobs}. 
The approximation relation is given by $X \bhord \tree$ iff 
for all $\ftr\in X$ there is a path in $\tree$ starting from the root labelled by $\ftr$. 
\end{example}

In the following, let us fix an observation structure \ple{\Obs,\obsord,\bhord} over $\BB$. 
Monitoring a property means trying to establish whether the system satisfies or violates it just relying on a (finite) observation of it. 
Formally, a monitor is a function \fun{\mnt}{\Obs}{\Vtre} where $\Vtre$ is the set of \emph{verdicts} $\set{\yv,\nv,\uv}$. 
This means that a monitor, given an observation, can either accept it (verdict $\yv$), or reject it (verdict $\nv$), or be inconclusive (verdict $\uv$). 
On $\Vtre$ it is clearly defined an \emph{information order} $\preceq$ given by 
$\uv\preceq\yv$ and $\uv\preceq\nv$. 
A monitor $\mnt$ is \emph{impartial} \cite{DongLS08} if it is monotone, that is, 
$\obsa\obsord\obsb$ implies $\mnt(\obsa) \preceq \mnt(\obsb)$, for all $\obsa\obsb\in\Obs$.
This means that it cannot retract conclusive verdicts  when considering finer observations. 

Given a property $\PP$, we define the \emph{abstract monitor} of $\PP$ to be a function 
\fun{\amnt\PP}{\Obs}{\Vtre} given as follows: 
\vspace*{-0.3cm}
\[\amnt\PP(\obsa) = \begin{cases}
\yv & \BB(\obsa)\subseteq \PP \\
\nv & \BB(\obsa)\cap\PP = \emptyset \\ 
\uv & \text{otherwise} 
\end{cases}\]
In other words, 
$\amnt\PP(\obsa) = \yv$ when all behaviours approximated by $\obsa$ \emph{satisfy} $\PP$, in which case we say that $\obsa$ \emph{positively determines $\PP$}, 
$\amnt\PP(\ftr) = \nv$ when all behaviours approximated by $\obsa$ \emph{violate} $\PP$, in which case we say that $\obsa$ \emph{negatively determines $\PP$}, and 
$\amnt\PP(\obsa) = \uv$ when no conclusive verdict can be emitted yet. 
We say that $\obsa$ \emph{determines $\PP$} when $\amnt\PP(\obsa)\ne\uv$. 
It is easy to check that the abstract monitor is impartial. 

\begin{proposition} \label{prop:mnt-mon} 
Let $\PP$ be a property on $\BB$. 
Then, for all $\obsa,\obsb\in\Obs$, 
$\obsa\obsord\obsb$ implies 
$\amnt\PP(\obsa)\preceq\amnt\PP(\obsb)$. 
\end{proposition} 

We say that an impartial monitor $\mnt$ is \emph{sound} for a property $\PP$ if, for all $\obsa\in\Obs$, 
$\mnt(\obsa) \preceq \amnt\PP(\obsa)$, that is, it cannot contraddict the abstract monitor of $\PP$. 
Moreover, we say that $\mnt$ is \emph{almost complete} if, for all $\obsa\in\Obs$, there exists $\obsb\in\Obs$ such that 
$\obsa\obsord\obsb$ and $\mnt(\obsb) = \amnt\PP(\obsa)$. 
In other words, $\mnt$ is almost complete when it can always eventually meet the verdict of the abstract monitor of $\PP$. 

The following lemmas relate the abstract monitor with set inclusion and complementation; where the following holds: 
$\dual\yv = \nv$, $\dual\nv = \yv$ and $\dual\uv = \uv$. 

\begin{lemma}\label{lem:mon-det}
Let $\PP,\PQ$ be properties on $\BB$ such that $\PP\subseteq\PQ$ and $\obsa\in\Obs$ be an observation.
Then, 
$\amnt\PP(\obsa) = \yv$ implies $\amnt\PQ(\obsa) = \yv$ and 
$\amnt\PQ(\obsa) = \nv$ implies $\amnt\PP(\obsa) = \nv$.
\end{lemma}

\begin{lemma} \label{lem:mnt-compl} 
Let $\PP$ be a property on $\BB$ and $\obsa\in\Obs$ be an observation.
Then, $\amnt{\BB\setminus\PP}(\obsa) = \dual{\amnt\PP(\obsa)}$. 
\end{lemma}

Using the abstract monitor, we define when a property $\PP$ is \emph{monitorable} (w.r.t. the notion of $\forall$-monitorable~\cite{DBLP:conf/fm/PnueliZ06}). 
This happens when the abstract monitor has always the possibility of emitting a positive or negative verdict. 

\begin{definition}
\label{def:monitorable}
A property $\PP$ on $\BB$ is \emph{monitorable} if, 
for every observation $\obsa\in\Obs$, there exists $\obsb\in\Obs$ such that 
$\obsa\obsord\obsb$ and 
$\amnt\PP(\obsb)\ne\uv$. 
\end{definition}
\noindent Thus, every observation has a refinement that determines the property. 
Another feature of monitorable properties is that they are closed under complementation. 

\begin{proposition} \label{prop:monitorable-compl}
Let $\PP$ be a monitorable property on $\BB$. 
Then, $\BB\setminus\PP$ is monitorable. 
\end{proposition} 


\begin{example} 
	\label{ex:non-mon} 
	We give some examples of non-monitorable properties. 
	\begin{enumerate}
	\item\label{ex:non-mon:1}
	Let us consider the observation structure in \Cref{ex:ltobs}.
	For the sake of clarity we refer to the usual syntax and semantics of LTL (see \cite{Pnueli77}).
	We write $\lsem{\ff}$ for the semantics of the LTL formula $\ff$.
	Let $\Events = \set{a,b,c,d}$ and consider the formula 
	$\ff = (a \land \F b) \lor (c \land \G\F d)$. 
	The property $\lsem\ff\subseteq \Events^\omega$ is clearly non-monitorable since all the traces $c\ftr \in \Ftr$ have no 
	finite extensions determining it.
	\item\label{ex:non-mon:2}
	Let us consider the first setting of \Cref{ex:btobs} 
	where $\Events = \set{a,b,c}$. 
	Consider the property $\PP$ satisfied by all those trees not containing a path starting from the root with an event $a$. 
	This property is not monitorable as, for instance, the trace $bb$ cannot be extended to reach a conclusive verdict.
\end{enumerate}
\end{example}


\section{Partial monitoring via (co)safety completions} 
\label{sec:completions} 

When we want to monitor a property $\PP$, 
usually we first check whether it is monitorable, to ensure that the monitor will not run forever even when it will never reach any conclusive verdict. 
In this section, we propose a different approach, describing its semantic foundations: 
instead of ruling out non-monitorable properties, we show it is always possible to find monitorable (over/under) approximations preserving either negative or positive verdicts. 
Moreover, combining such approximations, we define a generalised abstract monitor which detects when it will not be able to provide any conclusive verdict. 

To develop our approximations, we rely on standard lattice-theoretic notions for which we refer to \cite{DaveyP02}. 
In particular, the set of all properties over $\BB$, that is, the powerset $\wp(\BB)$, is a complete lattice w.r.t. subset inclusion. 
Then, the key idea is to look for approximations of properties in the classes of safety and cosafety properties, which are well-behaved subsets of $\wp(\BB)$. 

Let us start with safety properties. 
Informally, a safety property is a property for which we can always finitely determine whether a behaviour violates it. 

\begin{definition}[Safety property] 
	\label{def:safety}
	A property $\PP$ on $\BB$ is a \emph{safety property} if 
	for every $\bha\notin\PP$ there is an observation $\obsa\bhord\bha$ that negatively determines it. 
\end{definition}

\begin{proposition}
	\label{def:safety2}
	$\PP$ is a safety property on $\BB$ if and only if 
	for every $\bha\in\BB$, $\BB(\obsa)\cap\PP\ne\emptyset$  for each observation $\obsa\bhord\bha$ implies $\bha\in\PP$. 
\end{proposition}

This notion is pretty standard \cite{DBLP:conf/fm/PnueliZ06} and usually safety properties come with a crucial feature: they are always monitorable. 
However, in our abstract setting, this is not true in general. 
For instance, \refItem{ex:non-mon}{2} shows a safety property which is not monitorable. 
The issue with this example is that observations do not have enough structure to support monitorability of safety properties. 
Indeed, if we consider the same property in the second observation structure of \Cref{ex:btobs}, it is monitorable. 
Intuitively, the problem is that, if we use finite traces to monitor branching time properties, the monitor can fall into a branch where there is no hope to reach a conclusive verdict. 
Hence, to recover monitorability of safety properties, we need to assume a richer structure on observations (see below). 

\begin{definition}  \label{def:direct-obs}
An observation structure \ple{\Obs,\obsord,\bhord}  over $\BB$ is \emph{directed} if, 
for all $\obsa,\obsb\in\Obs$ and $\bha\in\BB$, $\obsa\bhord\bha$ and $\obsb\bhord\bha$ imply that there is $\obsc\in\Obs$ such that 
$\obsc\bhord\bha$ and $\obsa\obsord\obsc$ and $\obsb\obsord\obsc$. 
\end{definition}
\noindent In other words, the set $\Obs(\bha)$ is a directed set w.r.t. $\obsord$ for every $\bha\in\BB$. 
Note that, all the examples of observation structures we have seen are directed except for the first one in \Cref{ex:btobs}. 
Now, for directed observation structures, we can prove that safety properties are monitorable. 

\begin{proposition}  \label{prop:monitorable-sf} 
Let \ple{\Obs,\obsord,\bhord} be a directed observation structure on $\BB$. 
Then, every safety property is monitorable. 
\end{proposition} 

From now on let us assume a directed observation structure \ple{\Obs,\obsord,\bhord} over $\BB$.
Denote by $\SF$ the set of all safety properties over $\BB$. 
Then, the set $\SF$ is a suborder of $\wp(\BB)$ and the following  lemma  shows it is a closure system, that is, it is closed under arbitrary intersections. 

\begin{lemma}\label{lem:sf-cl} 
$\SF$ is a closure system. 
\end{lemma}

Like any closure system, $\SF$ induces a closure operator on $\wp(\BB)$:

\centerline{
$\clop[\SF](\PP) = \bigcap \{\PQ \in \SF \mid \PP\subseteq \PQ \}$
}
Being a closure operator, $\clop[\SF]$ is a monotone function on $\wp(\BB)$ w.r.t. subset inclusion and has the following two properties for any $\PP\subseteq\BB$: 
$\PP \subseteq\clop[\SF](\PP)$ and $\clop[\SF](\clop[\SF](\PP)) \subseteq \clop[\SF](\PP)$. 
In other words, $\clop[\SF](\PP)$ is \emph{the best safety (over)approximation of $\PP$} and we call it the \emph{safety completion of $\PP$}. 

We now give an alternative characterisation of $\clop[\SF]$. 
Let $\PP$ be a property on $\BB$. 
Define the set $\NR(\PP)$ of \emph{$\PP$-unrefutable} behaviours as follows: 

\centerline{
$\NR(\PP) = \set{ \bha\in\BB \mid \BB(\obsa)\cap\PP\ne\emptyset \text{ for all } \obsa\bhord\bha }$
}
\begin{proposition}
$\NR$ is a closure operator. 
\end{proposition}

Note that, by \cref{def:safety2}, $\PP$ is a safety property if and only if $\NR(\PP) \subseteq\PP$ if and only if $\PP$ is a fixpoint of $\NR$, as $\NR$ is a closure operator. 
Then, we get the following result. 

\begin{corollary}\label{cor:sf-alt} 
$\clop[\SF](\PP) = \NR(\PP)$, for all $\PP \subseteq \BB$. 
\end{corollary}

The key property of the safety completion of $\PP$ is that it exactly preserves all negative verdicts of $\PP$ as proved by the following theorem. 

\begin{theorem}  \label{thm:sf-nv} 
Let $\PP$ be a property on $\BB$ and $\obsa\in\Obs$. 
Then, $\amnt{\PP}(\obsa) = \nv$ if and only if $\amnt{\clop[\SF](\PP)}(\obsa) = \nv$. 
\end{theorem}

Therefore, if we are interested in detecting just violations of $\PP$, we can monitor its safety completion, which is necessarily monitorable and emits all and only negative verdicts of $\PP$. 
Moreover, when $\amnt{\clop[\SF](\PP)}(\obsa) = \yv$, we are sure that $\amnt\PP(\obsa)\ne \nv$ and, since an abstract monitor is impartial, we can safely stop the execution, as we will never receive a negative verdict. 

We now dualise everything considering cosafety properties. 
Informally, a cosafety property is a property for which we can always finitely determine whether a behaviour satisfies it. 

\begin{definition}[Cosafety property] 
A property $\PP$ on $\BB$ is a \emph{cosafety property} if 
for every $\bha\in\PP$, there exists an observation $\obsa\bhord\bha$ positively determining $\PP$. 
\end{definition}

\begin{proposition} \label{prop:co-safety} 
$\PP$ is a cosafety property if and only if $\BB\setminus\PP$ is a safety property. 
\end{proposition}

This proposition allows us to dualise all results obtained for safety properties. 
Denote by $\coSF$ the set of all cosafety properties on $\BB$.  
From \cref{prop:monitorable-compl,prop:monitorable-sf} we get that cosafety properties are monitorable and from \cref{lem:sf-cl} we get that $\coSF$, as a suborder of $\wp(\BB)$, is an interior system, that is, it is closed under arbitrary unions. 
Therefore, we get an interior operator 
on $\wp(\BB)$: 

\centerline{
$\inop[\coSF](\PP) = \bigcup \set{\PQ \in \coSF \mid \PQ \subseteq \PP }
$
}
Being an interior  operator, $\inop[\coSF]$ is a monotone function on $\wp(\BB)$ with respect to set inclusion and has the following two properties for any $\PP\subseteq\BB$
$\inop[\coSF](\PP) \subseteq \PP$ and $\inop[\coSF](\PP) \subseteq \inop[\coSF](\inop[\coSF](\PP))$. 
In other words, $\inop[\coSF](\PP)$ is \emph{the best cosafety (under)approximation of $\PP$} and we call it the \emph{cosafety completion of $\PP$}. 
%
%
%

It is easy to see that 
$\BB\setminus\clop[\SF](\PP) = \inop[\coSF](\BB\setminus\PP)$. 
Hence, the cosafety completion of $\PP$ exactly preserves positive verdicts of $\PP$, as proved below. 

\begin{corollary}  \label{cor:cosf-yv} 
Let $\PP$ be a property on $\BB$ and $\obsa \in \Obs$. 
Then, $\amnt\PP(\obsa) = \yv$ if and only if $\amnt{\inop[\coSF](\PP)}(\obsa) = \yv$. 
\end{corollary} 

Therefore, if we are interested in detecting just satisfaction of $\PP$, we can monitor its cosafety completion, which is necessarily monitorable and emits all and only positive verdicts of $\PP$. 
Moreover, when $\amnt{\inop[\coSF](\PP)}(\ftr) = \nv$, we are sure that $\amnt\PP(\ftr)\ne \yv$ and, since an abstract monitor is impartial, we can safely stop the execution as we will never receive a positive verdict. 

In summary, for any property $\PP$ on $\BB$, 
there are two monitorable properties 
$\inop[\coSF](\PP) \subseteq \PP \subseteq \clop[\SF](\PP)$, 
where the former captures exactly positive verdicts of $\PP$ and the latter captures exactly negative verdicts of $\PP$. 
However, combining the two completions, we get even more information: we can characterise the so called \emph{ugly} observations, that is, those observations after which it is not possible to emit any conclusive verdict. 
These are precisely those observations that the monitorability constraint rules out, but, if they can be detected, there is no need to restrict ourselves to monitorable properties. 

\begin{proposition} \label{prop:ugly}
Let $\PP$ be a property on $\BB$ and $\obsa\in\Obs$ an observation. 
Then, $\amnt{\clop[\SF](\PP)}(\obsa) = \yv$ and $\amnt{\inop[\coSF](\PP)}(\obsa) = \nv$ 
if and only if, 
for every $\obsb\in\Obs$ such that $\obsa\obsord\obsb$, $\amnt\PP(\obsb) = \uv$. 
\end{proposition} 

Relying on these results, we define a notion of generalised monitor as a function \fun{\gmnt}{\Obs}{\Vsei}, where $\Vsei$ is a set of verdicts consisting of six elements: 
$\yv$, $\nv$ and $\uv$ as before and $\uyv$, when no negative verdict can be reached, 
$\unv$, when no positive verdict can be reached, and 
$\nmv$, informally called \emph{giveup}, when no verdict at all can be reached. 
The information order $\preceq$ on $\Vsei$ is depicted in \cref{fig:inf-ord-v6}. 
\begin{figure}[t]
\centering
\framebox[0.55\textwidth]{
\begin{math}
\xymatrix{
\yv && \nmv && \nv \\ 
& \ar[lu] \uyv \ar[ru] && \ar[lu] \unv \ar[ru] \\ 
&& \ar[lu] \uv \ar[ru] && 
}
\end{math}
}
\caption{The information order on $\Vsei = \set{\yv,\nv,\uv,\uyv,\unv,\nmv}$.}
\label{fig:inf-ord-v6} 
\end{figure} 

Then, for any property $\PP$, we can define a \emph{generalised abstract monitor} to be a function 
\fun{\gamnt\PP}{\Obs}{\Vsei} given as follows: 
\vspace*{-0.3cm}
\[\gamnt\PP(\obsa) = \begin{cases}
\yv  & \amnt{\inop[\coSF](\PP)}(\obsa) = \yv \\ 
\nv  & \amnt{\clop[\SF](\PP)}(\obsa) = \nv \\ 
\nmv & \amnt{\clop[\SF](\PP)}(\obsa) = \yv \text{ and } \amnt{\inop[\coSF](\PP)}(\obsa) = \nv \\ 
\uyv & \amnt{\clop[\SF](\PP)}(\obsa) = \yv \text{ and } \amnt{\inop[\coSF](\PP)}(\obsa) = \uv \\ 
\unv & \amnt{\clop[\SF](\PP)}(\obsa) = \uv \text{ and } \amnt{\inop[\coSF](\PP)}(\obsa) = \nv \\ 
\uv  & \amnt{\clop[\SF](\PP)}(\obsa) = \uv \text{ and } \amnt{\inop[\coSF](\PP)}(\obsa) = \uv 
\end{cases}\] 
Note that cases not mentioned above are impossible due to \cref{lem:mon-det}. 
It is also easy to see that the generalised abstract monitor is impartial, meaning that 
$\gamnt\PP(\obsa) \preceq \gamnt\PP(\obsb)$, for all $\obsa,\obsb\in\Obs$ such that $\obsa\obsord\obsb$. 
Moreover, it can always eventually reach a conclusive verdict as proved by the following  theorem.

\begin{theorem}
Let $\PP$ be a property on $\BB$. 
Then, for every $\obsa\in\Obs$, there is $\obsb\in\Obs$  such that $\obsa\obsord\obsb$ and 
$\gamnt\PP(\obsb) \in \set{\yv,\nv,\nmv}$. 
\end{theorem}

In conclusion, to produce all verdicts of the generalised abstract monitor of a property $\PP$, we have just to look at two monitorable properties: its safety and cosafety completions. 
But we can go further: since the complement of a cosafety property is a safety property (\cref{prop:co-safety}), 
we can restrict our attention to just safety properties, by monitoring $\clop[\SF](\PP)$ and $\BB\setminus\inop[\coSF](\PP) = \clop[\SF](\BB\setminus\PP)$, and appropriately inverting verdicts. 
This is the strategy adopted in the following sections where we instantiate this technique in different monitoring settings. 

\begin{example} 
	\label{example_compl}
	Consider the property $\ff = (a \land \F b) \lor (c \land \G\F d)$ from \refItem{ex:non-mon}{1}. 
	Its safety completion is given by $\ff_1 = a \lor c$, while its cosafety completion by $\ff_2 = a\land \F b$. 
	We report below some verdicts of the generalised abstract monitor of $\ff$ obtained combining $\ff_1$ and $\ff_2$.  \\ 
	\centerline{$ 
	\begin{array}{cccc} 
		\ftr  &\quad \amnt{\lsem{\ff_1}}(\ftr) \quad &\quad \amnt{\lsem{\ff_{2}}}(\ftr) \quad &\quad \gamnt{\lsem{\ff}}(\ftr) \\ 
		\hline
		c\dots        & \yv & \nv & \nmv  \\ 
		a             & \yv & \uv & \uyv  \\ 
		b\dots        & \nv & \uv & \nv   \\ 
		a\dots b\dots & \yv & \yv & \yv
	\end{array}
$} 
\end{example}

\begin{remark}[Never Gonna Give You Up]
Note that, the generalised abstract monitor allows us to recognise non-monitorable properties. 
Indeed, by \cref{prop:ugly}, a property $\PP$ is monitorable if and only if $\gamnt\PP(\obsa)\ne \nmv$ for every $\obsa\in\Obs$.  
\end{remark}

\section{Partial Monitoring of Linear Time Properties}

In this section we present two ways of applying the general framework introduced in \Cref{sec:completions}.
First, we consider Temporal Logic, which is one of the most used formalism in RV, showing how we can synthesise a generalised monitor for them. 
Then, we focus on a formalism where only safety properties can be expressed, the Linear Time $\nu$-calculus, 
showing how it can be used to describe generalised monitors and proving it suffices to express those of all LTL properties. 


\subsection{Linear Temporal Logic}
\label{sec:ltl}

\begin{figure}[t]
	\centering
	\framebox[\textwidth]{
	\begin{math}
	\xymatrix @C=0.3em @R=0.3em{ 
	{Input} & {Formula} & {NBA} & & {NFA} & {DFA} & {\textrm{Moore machine}} \\
	& {\ff} \ar[r] & {\NBA_{\ff}} \ar[r] & {\FairA{\NBA_{\ff}}} \ar[r] 
	& {\NFA_{\ff}} \ar[r] & {\DFA_{\ff}} \ar[rd] & \\
	{\ff} \ar[ru] \ar[rd] & & & & & & {\mnt_\ff} \\
	& {\lnot \ff} \ar[r] & {\NBA_{\lnot \ff}} \ar[r] & {\FairA{\NBA_{\lnot \ff}}} \ar[r] 
	& {\NFA_{\lnot \ff}} \ar[r] & {\DFA_{\lnot \ff}} \ar[ru] & \\
	}
	\end{math}
	}
\caption{
NBA is Non-deterministic \Buchi Automaton, 
NFA is Non-deterministic Finite Automaton
and DFA is Deterministic Finite Automaton.}
\label{fig:fsm-steps}
\end{figure}
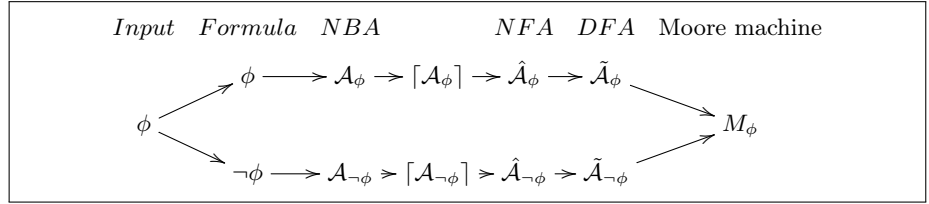

LTL is a commonly used specification language in RV. 
We use $\ff,\fp$ to range over LTL formulas and denote by $\lsem\ff$ the semantics of $\ff$, namely the set of infinite traces satisfying $\ff$. 
We refer to the literature \cite{DBLP:conf/focs/Pnueli77} for formal definitions (we report them in \cref{sec:app-ltl} for reader's convenience). 
Given $\ff$, \Cref{fig:fsm-steps} shows the canonical procedure for synthesising its monitor~\cite{10.1145/2000799.2000800}.

Informally, we start from $\ff$ and 
first we synthesize the equivalent non deterministic Büchi automaton (NBA) $\NBA_{\ff}$. 
Then, we set as \emph{final} all states whose
accepted language is not empty (\ie the states that can reach infinitely often a final one), 
obtaining the NBA $\FairA{\NBA_{\ff}}$. 
After that, we treat such automaton as a NFA and we \emph{determinise} it. 
We do the same process for $\neg\ff$ and, finally, 
we combine the DFAs $\DFA_{\ff}$ and $\DFA_{\lnot \ff}$ to obtain the monitor.

\begin{figure}[t]
	\centering
	\framebox[\textwidth]{
	\begin{math}
	\xymatrix @C=0.3em @R=0.3em{ 
	{Input} & & {NBA} & & {NFA} & {DFA} & {\bf Completions} & {\bf Gen. Mon.} \\
	& {\ff} \ar[r] & {\NBA_{\ff}} \ar[r] & {\FairA{\NBA_{\ff}}} \ar[r] 
	& {\NFA_{\ff}} \ar[r] & {\DFA_{\ff}} \ar[r] & {\mnt_{\clop[\SF](\ff)}} \ar[rd] & \\
	{\ff} \ar[ru] \ar[rd] & & & & & & & {\gmnt_\ff} \\
	& {\lnot \ff} \ar[r] & {\NBA_{\lnot \ff}} \ar[r] & {\FairA{\NBA_{\lnot \ff}}} \ar[r] 
	& {\NFA_{\lnot \ff}} \ar[r] & {\DFA_{\lnot \ff}} \ar[r] & {\mnt_{\clop[\SF](\neg\ff)}} \ar[ru] &\\
	}
	\end{math}
	}
\caption{Steps required to generate a generalized monitor for a LTL property $\ff$.}
\label{fig:new-steps}
\end{figure}
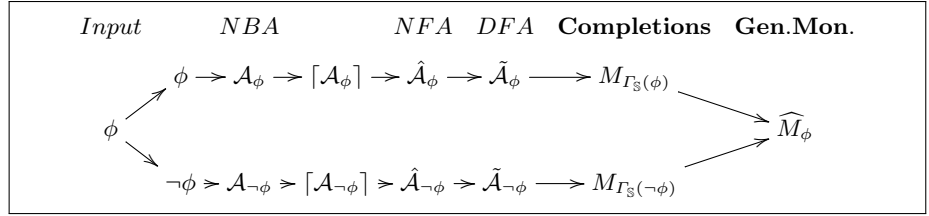

\Cref{fig:new-steps} shows how we can refine the steps in \Cref{fig:fsm-steps}
to build a generalized monitor for $\ff$.
Notably, the difference with respect to \Cref{fig:fsm-steps} lies in the last steps.
Indeed, we transform the DFA $\DFA_{\ff}$ into a monitor $\mnt_{\DFA_{\ff}}$, assigning the verdict $\yv$ to those states that can reach \emph{only} final states. 
Denote by $\Language\Automaton$ the language accepted by an automaton $\Automaton$.
It is easy to check that 
$\mnt_{\DFA_{\ff}}$ coincides with the abstract monitor of $\Language{\FairA{\NBA_\ff}}$, as proved below. 
\begin{lemma}
	For every formula $\ff$ and finite trace $\ftr$, $\mnt_{\DFA_{\ff}}(\ftr) = \amnt{\Language{\FairA{\NBA_{\ff}}}}(\ftr)$. 
\end{lemma}
Then, we observe that the third step in \Cref{fig:new-steps} corresponds to a safety completion.
\begin{lemma}
	Let $\NBA$ be a NBA. Then $\Language{\FairA\NBA} = \clop[\SF](\Language\NBA)$.
\end{lemma}  
This shows we have constructed the abstract monitors for the safety completions $\clop[\SF](\lsem{\ff})$ and $\clop[\SF](\lsem{\neg\ff})$, thus combining them, as described at the end of \cref{sec:completions}, we obtain the generalised abstract monitor of $\ff$. 
\begin{theorem}
	For every formula $\ff$ and finite trace $\ftr$, $\gmnt_{\ff}(\ftr) = \gamnt{\lsem{\ff}}(\ftr)$. 
\end{theorem} 

\begin{figure}[!htb]
    \centering
    \begin{minipage}{.33\textwidth}
        \centering
        \includegraphics[width=1.0\linewidth]{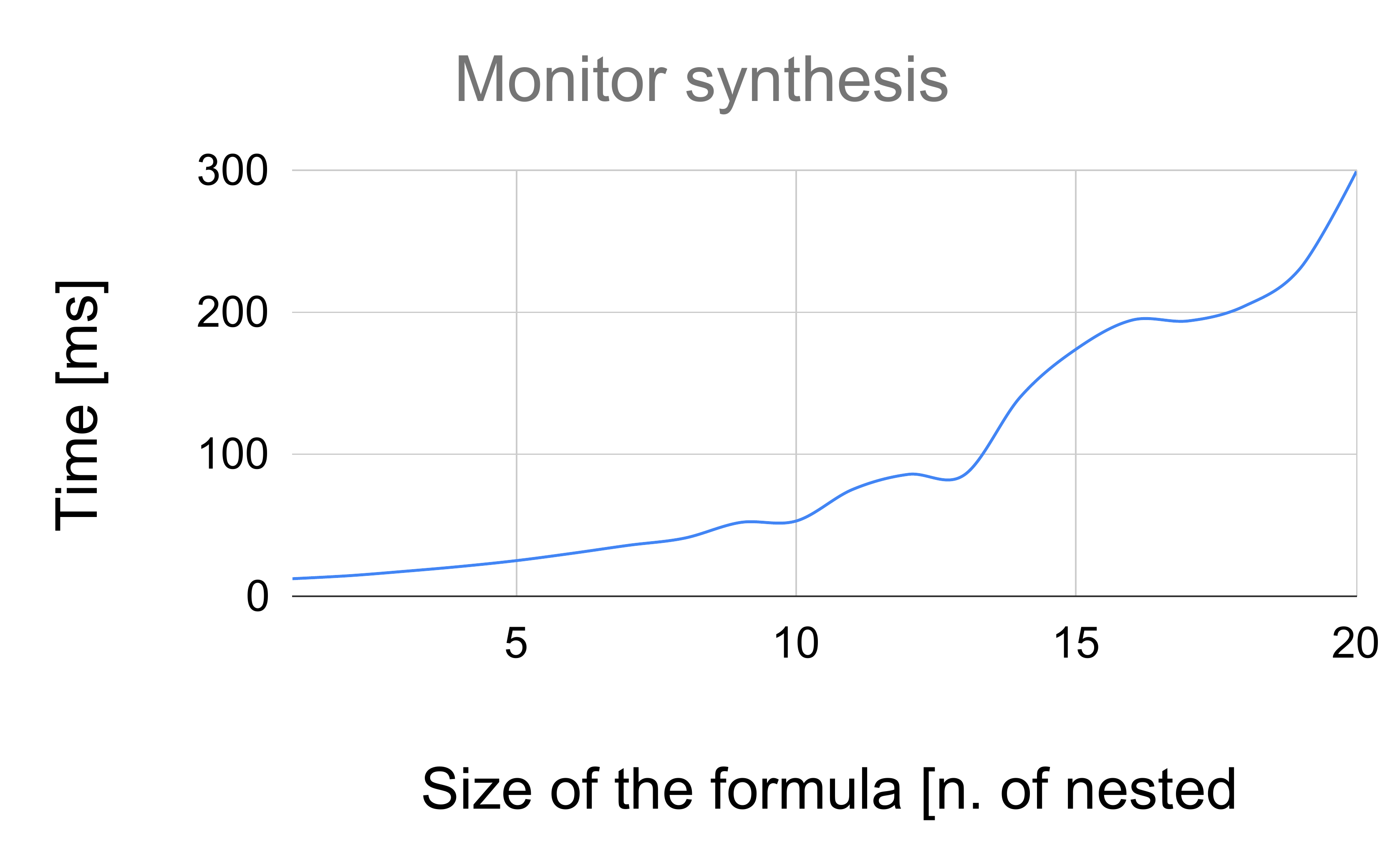}
        \caption{Monitor synthesis.}
        \label{fig:ltl_generation}
    \end{minipage}%
    \begin{minipage}{.33\textwidth}
        \centering
        \includegraphics[width=1.0\linewidth]{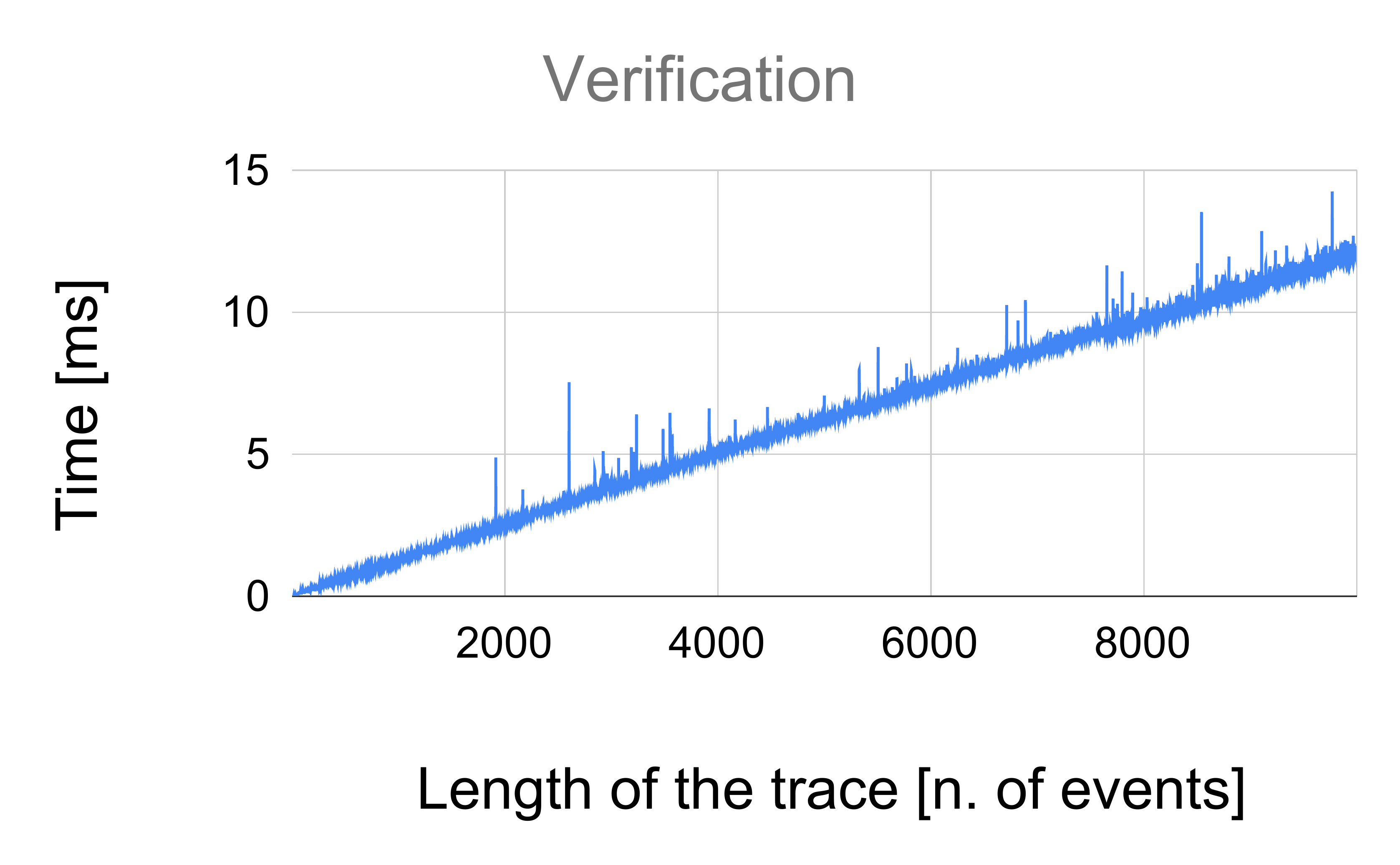}
        \caption{Verification.}
        \label{fig:ltl_verification}
    \end{minipage}
    \begin{minipage}{.33\textwidth}
        \centering
        \includegraphics[width=1.0\linewidth]{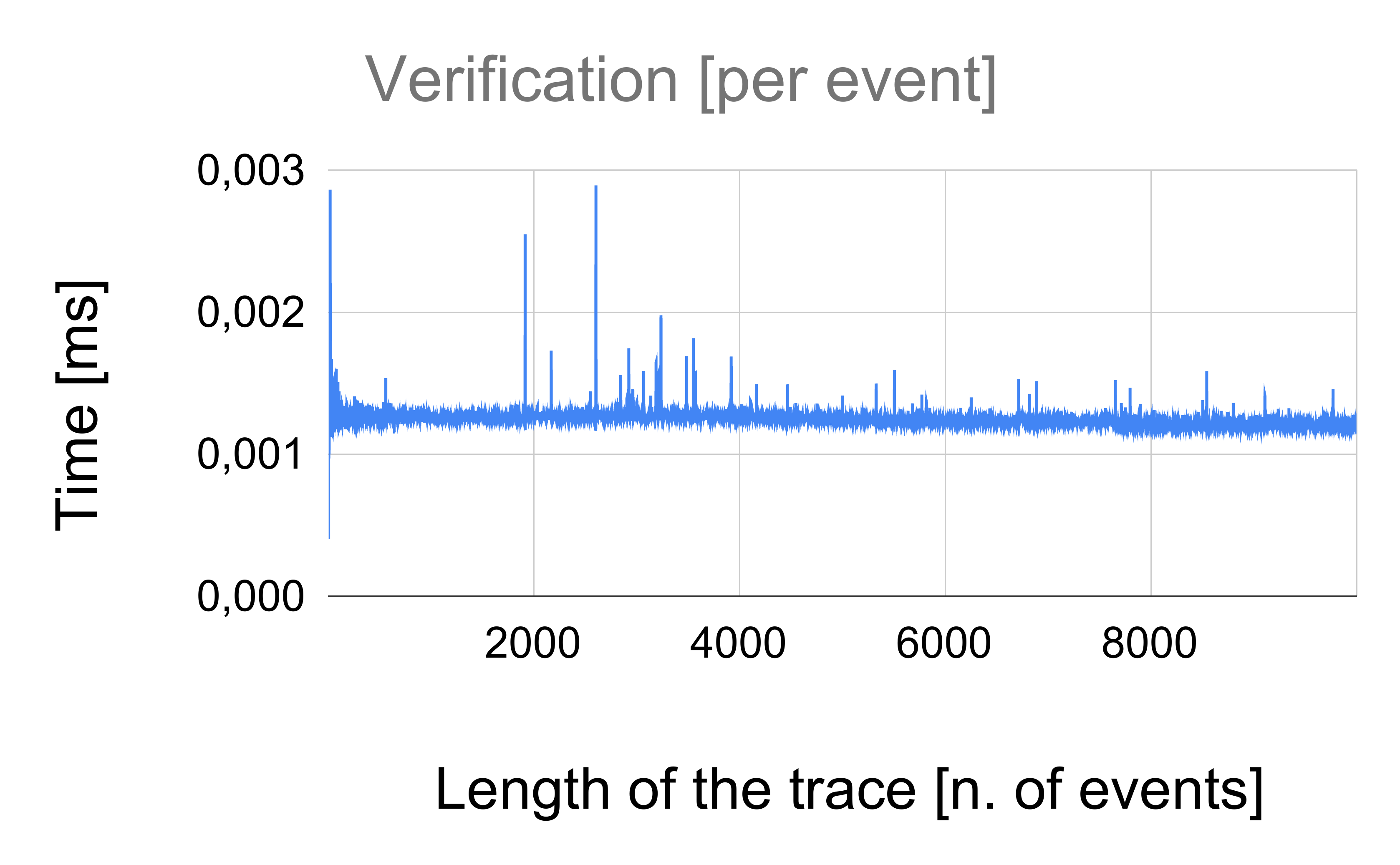}
        \caption{Verification (event).}
        \label{fig:ltl_verification_per_event}
    \end{minipage}
\end{figure}

\paragraph{Implementation and Experiments.}
A prototype implementing the LTL instantiation of our general framework is available as supplementary material. 
In more detail, the tool is implemented in Python exploiting the Spot library\footnote{\scriptsize\url{https://spot.lrde.epita.fr/}}~\cite{DBLP:conf/mascots/Duret-LutzP04}, to handle the translation from LTL to B\"{u}chi Automata, and the Automata library\footnote{\scriptsize\url{https://github.com/caleb531/automata}}, to handle the subsequent transformations (to NFA and DFA). 
We tested our tool on the example properties presented in this paper; then, we carried out experiments to validate it and analyse its performance. The results are reported in \cref{fig:ltl_generation,fig:ltl_verification,fig:ltl_verification_per_event}.

\cref{fig:ltl_generation} reports the time required to synthesise a monitor given an LTL formula. As in the standard LTL generation, the complexity is exponential w.r.t. the size of the formula. However, more importantly for runtime verification, the execution time required for the actual verification is still linear (\cref{fig:ltl_verification}). 
Indeed, the per event verification is constant\footnote{This is good from an overhead perspective assuming the monitor is used incrementally to analyse the system while the latter is running} and stabilises around $0.001$ [ms]. Note that, the experiments have been carried out on a large set of randomly generated LTL formulas and traces of events (more than 100000 formulas).


\subsection{Generalised Monitoring via Linear Time $\nu$-calculus}
\label{sec:monitor-nu}

As we have seen at the end of \cref{sec:completions}, safety properties suffices to describe generalised monitors. 
Following this idea, in this section we consider a language where only safety properties
can be expressed, 
the Linear Time $\nu$-calculus, 
showing how it can be used to describe generalised monitors and proving it can express those of all LTL properties. 
That is, this language is enough to specify monitors of a much more expressive logic. 

Let us assume a countable set $\AP$ of \emph{atomic propositions} ranged over by $\pp,\pq$, together with an interpretation function 
\fun{\psem{-}}{\AP}{\wp(\Events)}, interpreting each atomic proposition as a set of events. 
Let $\itr \in \Itr$, we denote by 
$\itr_i$ the $i$-th element of $\itr$ and by 
$\itr_{\geq i}$ the trace starting from the $i$-th element of $\itr$.
\Cref{fig:cstx_csem} shows syntax and semantics of such a calculus, 
a purely coinductive (\ie without $\mu$ operator) fragment of the Linear Time $\mu$-calculus \cite{BarringerKP86,DBLP:conf/popl/Vardi88,DaxHL06} that we dub LT$\nu$. 
Notably, the syntax is inductively defined while the semantics coinductively.
We set $\csem\tt = \set{\itr \in \Itr \mid w \cmods \tt}$.

\begin{figure}[t]
\centering
\framebox[\textwidth]{
\begin{mathpar}
	\tt,\ts \Coloneqq \top \mid \bot 
		\mid \pp \mid \co\pp \mid \tt\land\ts 
		\mid \tt\lor\ts \mid \Next\tt \mid X \mid \rec X \tt
	\\\\
	\inferrule[c-top]{\mathstrut}{\itr \cmods \top}
	\and
	\inferrule[c-prop]{\mathstrut}{\itr \cmods \pp} ~ \itr_0 \in \psem{\pp}
	\and
	\inferrule[c-coprop]{\mathstrut}{\itr \cmods \co\pp} ~ \itr_0 \not\in \psem{\pp}
	\and
	\inferrule[c-next]{\itr_{\geq 1} \cmods \tt}{\itr \cmods \Next\tt}
	\\
	\inferrule[c-or-l]{\itr \cmods \tt}{\itr \cmods \tt \lor \ts}
	\and
	\inferrule[c-or-r]{\itr \cmods \ts}{\itr \cmods \tt \lor \ts}
	\and
		\inferrule[c-and]{
		\itr \cmods \tt \\ \itr \cmods \ts
		}{
		\itr \cmods \tt \land \ts
		}
	\and
	\inferrule[c-rec]{\itr \cmods \tt\subst {\rec X \tt} X}{\itr \cmods \rec X \tt}
\end{mathpar}
}
\caption{Syntax and Semantics of Linear Time $\nu$-calculus.}
\label{fig:cstx_csem}
\end{figure}

\begin{example}
	We fix $\Events = \PFstar\AP$, the set of non-empty subsets of $\AP$. 
	Then, $\psem\pp = \set{\act \mid \pp \in \act}$.
	Let $\tt = \rec\var{\pp \land \Next\var}$. 
	Then, $\csem\tt = \set{\itr \mid \forall i.\pp \in \itr_i}$.
\end{example}

Without loss of generality, we assume that terms are \emph{contractive}, i.e., variables occur guarded by a next operator. 
Indeed, for every LT$\nu$ term, there is an equivalent contractive one \cite{DaxHL06}.
The next theorem states that the semantics of a term is a safety property.

\begin{proposition}
	\label{thm:csem_sf}
	Let $\tt$ be a LT$\nu$ term. Then $\csem\tt$ is a safety property.
\end{proposition}

The proof of \cref{thm:csem_sf} relies on \Cref{def:safety} which requires the 
characterization of a relation $\itr \not\cmods \tt$, meaning that $\itr \not\in \csem\tt$, 
which, being the dual of $\cmods$, is defined inductively. 

As usual, we can use LT$\nu$ terms as monitors, by 
inductively defining a labeled transition system  
$\tt \red\ev \ts$ (see \Cref{sec:app-ltnu-def}), 
meaning
that $\tt$ reduces to $\ts$ when the event $\ev$ occurs. 
We write $\tt \not\rightarrow$ when there are no $\ev, \ts$ such that $\tt \red\ev\ts$.
Let $\reds{}$ be the reflexive transitive closure of $\red{}$ 
and $\ireds{}$ be the coinductive predicate obtained by applying infinitely many times $\red{}$ 
(\ie $\tt \ireds\itr$ means that $\tt$ reduces infinitely many times using $\itr$). 

\begin{proposition}[Correctness of $\ireds{}$]
	\label{thm:ireds-correct}
	Let $\itr \in \Itr$ and $\tt$ an LT$\nu$ term. Then $\itr \cmods \tt$ iff $\tt \ireds\itr$.
\end{proposition}

Given a LT$\nu$ term $\tt$ we can build a monitor $\mnt_\tt: \Ftr \rightarrow \Vtre$ as follows: 
\vspace*{-0.3cm}
\[
\mnt_\tt(\ftr) = 
	\begin{cases}
		\yv & \tt \reds\ftr \ts$ \text{ and } $\prf\ts		\\
		\nv & \tt \reds\ftr \ts \text { implies } \ts \not\rightarrow  \\ 
		\uv & \text{otherwise} 
	\end{cases}
\]
The monitor returns $\nv$ when it gets stuck, while it returns $\yv$ when it reaches a term whose semantics is $\Itr$. 
To check this, we rely on a coinductive proof system (see \Cref{sec:app-ltnu-def}), which is a minimal variation of the one in \cite{DaxHL06} for the Linear Time $\mu$-calculus, satisfying 
$\csem\tt = \Itr$ iff $\prf\tt$, for every LT$\nu$ term $\tt$. 

According to \Cref{sec:completions}, in order to monitor an arbitrary property $\PP$, we have to
describe the monitors of its (co)safety completions. 
This can be done by specifying two safety properties $\PP_\SF$ and $\PP_\coSF$ such that $\PP_\SF\cup\PP_\coSF = \Itr$, as this is equivalent to $\Itr\setminus\PP_\coSF \subseteq \PP_\SF$ and $\Itr \setminus \PP_\coSF$ is a cosafety property. 
Therefore, 
given a pair of LT$\nu$ terms $\tt_{\SF}, \tt_{\coSF}$ such that $\csem{\tt_\SF}\cup\csem{\tt_\coSF} = \Itr$, 
which can be checked looking for a derivation of $\prf{\tt_\SF\lor\tt_\coSF}$, 
describing, respectively, the safety completion and the complement of the cosafety completion of some property, 
we obtain a generalised monitor for $\PP$ by combining verdicts of $\mnt_{\tt_\SF}$ and $\mnt_{\tt_\coSF}$. 

\begin{example}
	Let $\Events = \AP = \set{a,b,c,d}$ and $\psem\pp = \set\pp$.
	Consider again the LTL property $\ff = (a \land \F b) \lor (c \land \G\F d)$ from \Cref{ex:non-mon}.
	It requires to reject all traces not starting with $a$ or $c$ and 
	to accept all traces starting with $a$ and containing some $b$. 
	This can be specified by the following two LT$\nu$ terms: 
	$\tt_{\SF} = a \lor c$ and $\tt_{\coSF} = \co{a} \lor (\rec\var{\co{b} \land \Next\var})$, 
	remembering that accepted traces are those rejected by $\tt_\coSF$. 
	It is easy to see that $\csem{\tt_{\SF} \lor \tt_{\coSF}} = \Itr$.
\end{example}

Finally, the next theorem shows that by LT$\nu$ we can specify monitors for all LTL formulas. 

\begin{theorem}[Expressiveness w.r.t. LTL]
	\label{thm:ltl-ltnu}
	Let $\ff$ be a LTL formula. Then there exists $\tt_{\SF}$ such that $\clop[\SF](\lsem\ff) = \csem{\tt_{\SF}}$.
\end{theorem}

The proof of \Cref{thm:ltl-ltnu} is based on an algorithmic encoding of \Buchi automata as LT$\nu$ terms, computing the safety completion of the language accepted by the automaton. 
Such encoding can be applied to all those automata obtained from LTL formulae.

\begin{figure}[!htb]
    \centering
    \begin{minipage}{.5\textwidth}
        \centering
        \includegraphics[width=1.0\linewidth]{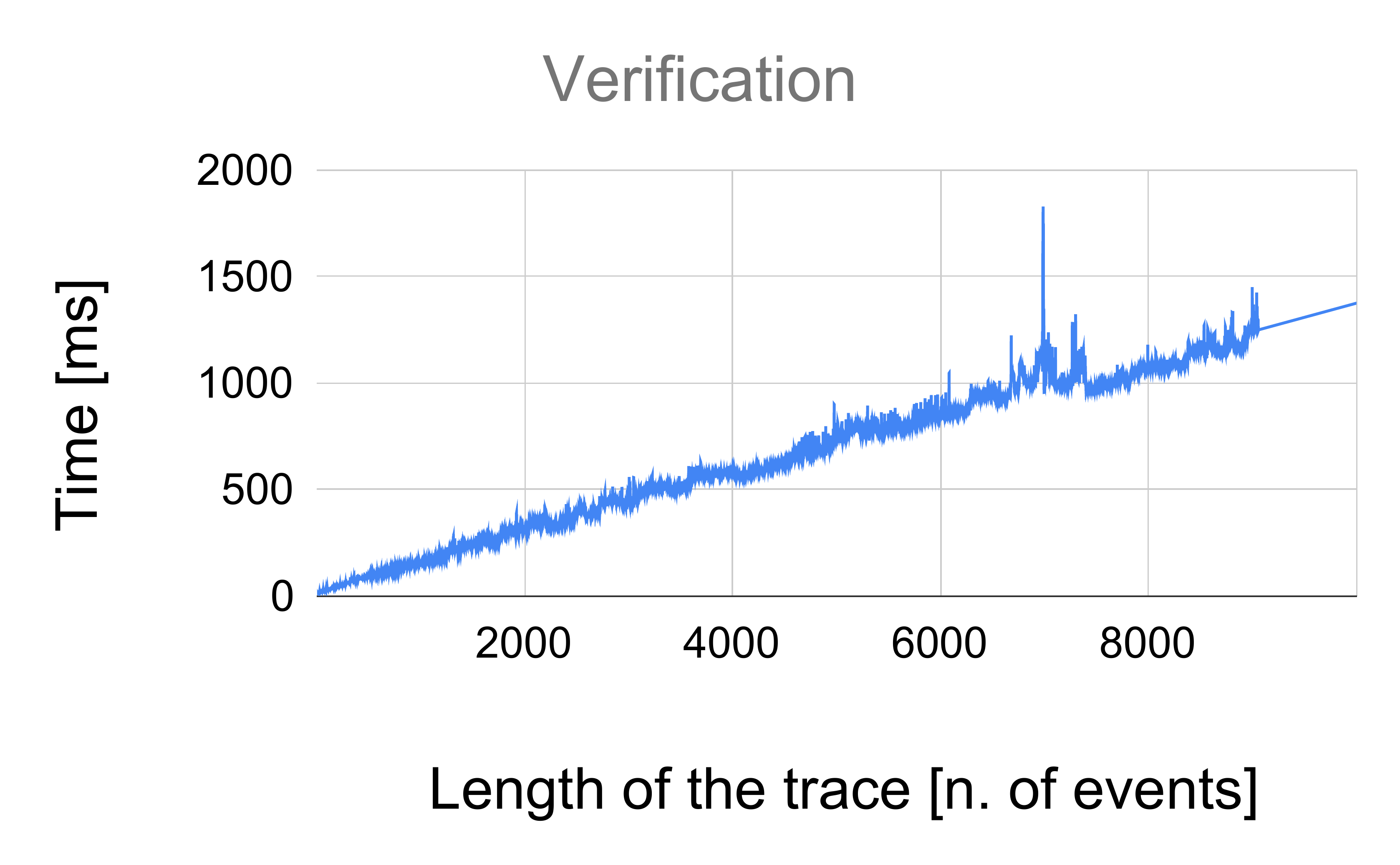}
        \caption{Verification.}
        \label{fig:nu_verification}
    \end{minipage}%
    \begin{minipage}{.5\textwidth}
        \centering
        \includegraphics[width=1.0\linewidth]{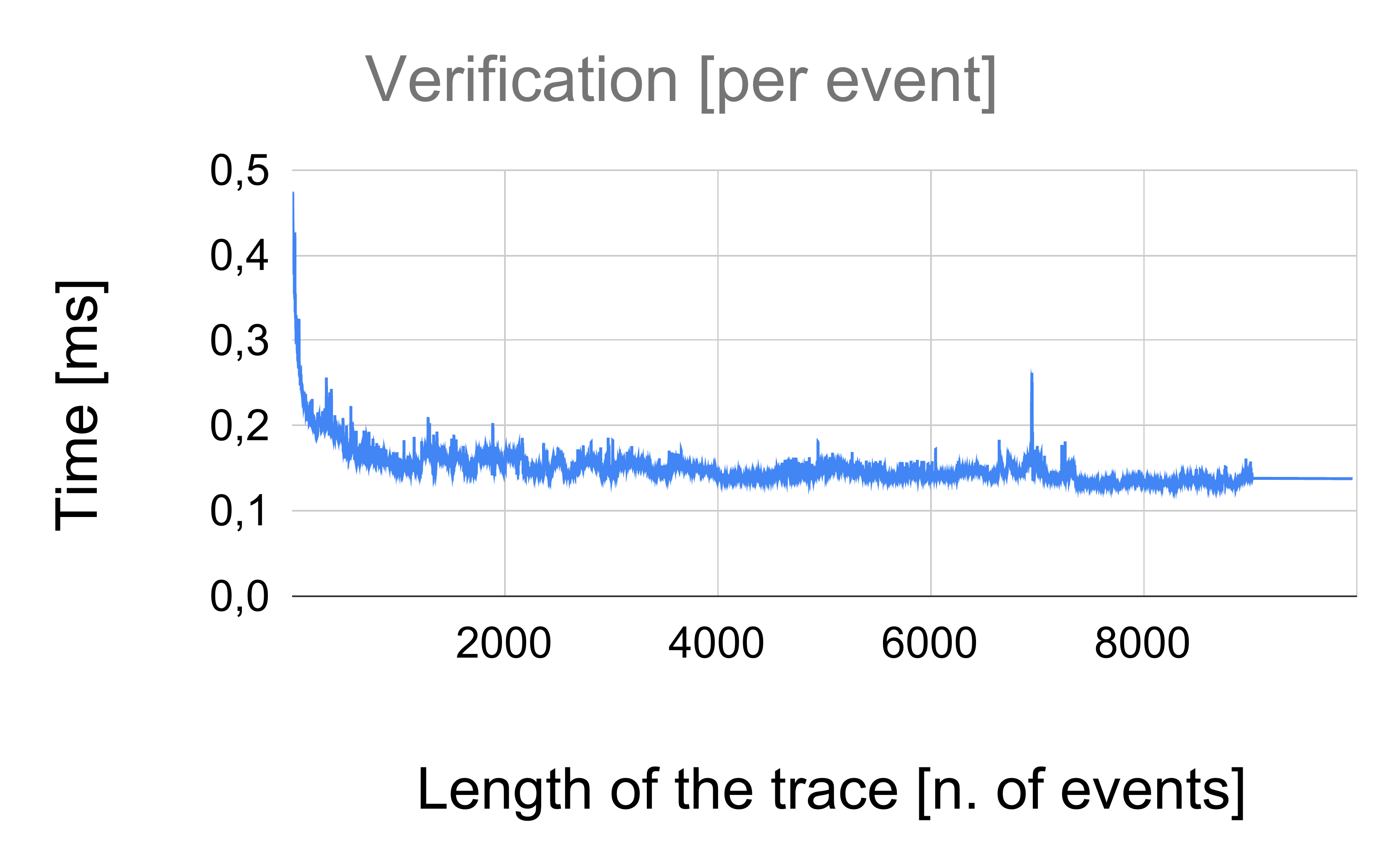}
        \caption{Verification (event).}
        \label{fig:nu_verification_per_event}
    \end{minipage}
\end{figure}

\paragraph{Implementation and Experiments.}
A prototype implementing the Linear Time $\nu$-calculus instantiation of our general framework is available  supplementary material. 
In more detail, the tool is implemented in SWI-Prolog. 
We chose SWI-Prolog due to its native support for coinduction and cyclic terms. 
Thanks to these features, the implementation of LT$\nu$ terms syntax and transition system is straightforward. 
As for the LTL prototype, we carried out experiments to validate our tool and analyse its performance as well. 
The obtained results are reported in \cref{fig:nu_verification,fig:nu_verification_per_event}.

Since we directly start from the LT$\nu$ terms, there is no synthesis step. 
\cref{fig:nu_verification} reports the execution time required to verify a given trace of events. As before, such execution is still performed in linear time w.r.t. the length of the analysed trace. Moreover, the incremental verification is still applicable as well, since the execution time per single event is maintained constant and stabilises around 0.15 [ms] (\cref{fig:nu_verification_per_event}). 
As before, the experiments have been carried out on a large set of randomly generated properties (LT$\nu$ terms in this case) and traces of events (more than 100000 randomly generated terms have been tested).

\section{Related Work}
\label{sec:related}

The present work can be seen as a foundational counterpart of~\cite{DBLP:journals/corr/abs-2110-12585}, where the authors present the idea of detecting ugly prefixes of LTL properties at runtime from a practical perspective. 
Indeed, they focus only on implementation aspects and 
no study on the theory behind it is done whatsoever. 
To provide such a foundation in a very general setting, covering both linear and branching time, we build on the framework in~\cite{StuckiSSB19}. 
They consider also gray-box monitoring and a concrete instantiation on hyperproperties, which we do not address, but they are both interesting directions for further work. 

In this paper, we consider the notion of monitorability introduced by Pnueli and Zaks~\cite{DBLP:conf/fm/PnueliZ06} and Peled and Havelund~\cite{DBLP:conf/birthday/PeledH18}, which generalises the first proposal~\cite{DBLP:journals/entcs/KimKLSV02}. 
This is one of the most commonly adopted in the literature and it has a more operational nature. 
That is, it focuses on the fact that a monitor should not run uselessly 
rather than on the verdicts it can produce. 

Other notions of monitorability are considered, for instance, by Aceto et al.~\cite{AcetoAFIL19popl}.  
Their completely monitorable properties corresponds to properties that are both safety and cosafety, while partially monitorable properties correspond to either safety or cosafety properties, depending on whether one focuses on satisfaction or violation. 
Hence, in the linear-time these notions are stronger than the one we consider, while in branching-time they are incomparable, since, as we have noticed, 
depending on the observation structure, (co)safety properties may not be monitorable in our setting. 
 
Aceto et al.~\cite{AcetoAFIL21} introduce the notion of \emph{best monitorable consequence}, which can be seen as an instance of our best safety approximation  of a property, at least when no additional condition on monitors is imposed. 
They also provide a syntactic construction to compute the best monitorable consequence of $\textsf{recHML}$ formulas, which could be a good starting point to obtain an instance of generalised monitors in a branching-time setting.\footnote{The observation structure they use is not directed (\cref{def:direct-obs}), hence we will need to adjust it, for instance considering hyperproperties.} 

On a more general side, different works can be found on the monitorability of LTL properties.
In~\cite{DBLP:conf/csl/TabuadaN16}, an extension of LTL called robust LTL (rLTL) is presented. With respect to LTL, rLTL semantics allows for distinguishing various ``degrees'' to which a
trace violates a property. This is obtained through five different verdicts (instead of the standard three), which quantitatively denote how much the property has been violated. It is important to note that in rLTL, all properties become monitorable~\cite{DBLP:conf/hybrid/MascleNSTW020}; simply because the resulting monitor does not look for the complete satisfaction (resp., violation) of the property, but it settles for a certain degree of satisfaction (resp., violation) of the latter.

In~\cite{DBLP:conf/cav/BartocciBNR18}, a two-step approach to address monitorability of LTL properties is proposed. Specifically, the satisfaction (resp., violation) of the LTL property is given by taking into consideration a prediction over the future of the analysed trace. Differently from other works~\cite{DBLP:conf/nfm/ZhangLD12}, such prediction is only based upon past experience, \textit{i.e.}, by only looking at how the trace behaved in the past. According to such prediction, the LTL property is evaluated considering how far the trace is supposed to be from satisfying or violating the property. Moreover, such distance is evaluated considering the safety and cosafety aspects of the property.  

In~\cite{DBLP:conf/rv/HenzingerS20}, an approach to extend the set of monitorable LTL properties through assumptions over the system is presented. Thanks to the additional information on the system, not all possible future continuations, given a prefix, are considered, but only a certain subset of them. Consequently, properties that are considered non-monitorable in general, can be monitored when such subset of traces reduces to only satisfying (resp., violating) traces.

With respect to the previously listed works, our contribution moves forward towards the concept of partial RV. Instead of focusing on extending LTL semantics, or presenting a predictive version of the latter, we present a general-purpose approach to partially monitor formal properties. Since we tackle this problem at a semantic level, our theoretical contribution is not limited to LTL; even though we present a possible instantiation exploiting LTL. Moreover, works in literature are more focused on estimating fully non-monitorable properties\footnote{Where with ``fully'' non-monitorable, we mean properties for which there are no prefixes that determine them neither positively, nor negatively.}, while we care for properties that are indeed non-monitorable in general, but can be monitored in some scenarios (\ie some prefixes can still determine the property at runtime). In fact, we present a solution that can be used both for monitoring the monitorable parts of a given property, and at the same time, to safely recognise when to stop the verification; since no final verdict will ever be concluded.

\section{Conclusions and Future Work}
\label{sec:conclusions-future-work}

This paper introduces partial RV,  studying its semantic foundations. 
We show how to build monitors capable to recognise when they will not be able to reach any conclusive verdict, and then to stop their execution. 
In this way, we can \emph{partially} monitor any property, avoiding issues related to non-monitorability, that is, having monitors running forever even though they will never be able to reach any conclusive verdict. 
We show that this can be achieved by combining safety and cosafety completions, which are well-behaved monitorable over/under approximations of any property. 
These results are developed independently from specific setups, covering both branching and linear time settings. 
Then, we present concrete instantiations of the framework for linear time properties, using LTL and Linear Time $\nu$-calculus as concrete languages. 
In both cases, we developed prototypes, available as supplementary material, with an empirical evaluation of their performances. 

As future work, we are planning to further test our prototypes on a set of well-known properties (a starting point could be the ones listed in~\cite{10.1145/2000799.2000800}).
On a more theoretical perspective, we are considering to further explore the branching time scenario, and its implications in our monitoring setting.


\bibliographystyle{splncs04}
\bibliography{main}

\newpage
\appendix
\section{Appendix}
\label{sec:appendix}

We detail the proofs of all the results.


\subsection{Supplement to \Cref{sec:completions}}
\begin{proposition}
Let \ple{\Obs,\obsord,\bhord} be a directed observation structure on $\BB$. 
Then, every safety property is monitorable. 
\end{proposition} 
\begin{proof}
Let $\PP\subseteq \BB$ be a safety property and $\obsa\in\Obs$  an observation. 
We distinguish two cases. 
If $\BB(\obsa)\subseteq\PP$, then $\amnt\PP(\obsa) = \yv$, as needed. 
Otherwise, there is $\bha\in\BB(\obsa)$ such that $\bha\notin\PP$. 
Since $\PP$ is safety, by definition, there is $\obsb\bhord\bha$ such that $\amnt\PP(\obsb) = \nv$. 
Since \ple{\Obs,\obsord,\bhord} is directed, there is $\obsc\in\Obs$ such that 
$\obsc\bhord\bha$ and $\obsa\obsord\obsc$ and $\obsb\obsord\obsc$. 
By impartiality of the abstract monitor, we get $\amnt\PP(\obsc) = \nv$ as needed. 
\end{proof}

\begin{lemma}
$\SF$ is a closure system. 
\end{lemma}
\begin{proof}
Let $X\subseteq \SF$ be a set of safety properties. 
We have to show that $\bigcap X \in \SF$. 
Consider $\bha\notin\bigcap X$, then by definition of intersection, there is a property $\PP \in X$ such that $\bha\notin\PP$. 
Since $X\subseteq\SF$, we have that $\PP$ is a safety property, hence, by definition, there is an observation $\obsa\bhord\bha$  such that $\BB(\obsa)\cap\PP = \emptyset$. 
Then, because $\PP\in X$, we have $\bigcap X\subseteq \PP$ and so $\BB(\obsa) \cap \left(\bigcap X\right) \subseteq \BB(\obsa) \cap\PP = \emptyset$, hence 
$\BB(\obsa)\cap\left(\bigcap X\right) = \emptyset$, that is, $\obsa$ negatively determines $\bigcap X$. 
This proves that $\bigcap X$ is a safety property, as needed. 
\end{proof}

\begin{proposition}
$\NR$ is a closure operator. 
\end{proposition}
\begin{proof}
It is easy to see that $NR$ is monotone ($\PP\subseteq\PQ$ implies $\NR(\PP)\subseteq\NR(\PQ)$) and extensive ($\PP\subseteq\NR(\PP)$). 
Hence, we have only to check that $\NR(\NR(\PP))\subseteq\NR(\PP)$. 
Let $\bha\in\NR(\NR(\PP))$, hence, for every observation $\obsa\bhord\bha$, there is $\bhb\in\\NR(\PP)$ such that $\obsa\bhord\bhb$. 
Thus, for every observation $\obsb\bhord\bhb$, there is $\bhc\in\PP$ such that $\obsb\bhord\bhc$. 
In particular, we get that, for every observation $\obsa\bhord\bha$, there exists $\bhc\in\PP$ such that $\obsa\bhord\bhc$, hence $\bha\in\NR(\PP)$, as needed. 
\end{proof}

\begin{corollary}
$\clop[\SF](\PP) = \NR(\PP)$, for all $\PP \subseteq \BB$. 
\end{corollary}
\begin{proof}
It is straightforward as two closure operators with the same set of fixpoints are equal. 
\end{proof}

\begin{theorem} 
Let $\PP$ be a property on $\BB$ and $\obsa\in\Obs$. 
Then, $\amnt{\PP}(\obsa) = \nv$ if and only if $\amnt{\clop[\SF](\PP)}(\obsa) = \nv$. 
\end{theorem} 
\begin{proof}
The right-to-left implication follows from \cref{lem:mon-det} as $\PP \subseteq \clop[\SF](\PP)$. 
Towards a proof of the other implication, assume $\BB(\obsa) \cap \PP = \emptyset$ and consider 
$\bha \in \BB(\obsa) \cap \clop[\SF](\PP)$. 
Then, $\obsa\bhord\bha$  and $\bha\in\clop[\SF](\PP)$. 
By \cref{cor:sf-alt}, we have $\clop[\SF](\PP) = \NR(\PP)$, hence we get that, for every observation $\obsb\bhord\bha$, $\BB(\obsb)\cap\PP\ne\emptyset$. 
Hence, in particular, since $\obsa\bhord\bha$, we get  $\BB(\obsa)\cap\PP\ne\emptyset$, which is a contradiction. 
As a consequence, we have $\BB(\obsa)\cap\clop[\SF](\PP) = \emptyset$, as needed. 
\end{proof}

\begin{corollary} 
Let $\PP$ be a property on $\BB$ and $\obsa \in \Obs$. 
Then, $\amnt\PP(\obsa) = \yv$ if and only if $\amnt{\inop[\coSF](\PP)}(\obsa) = \yv$. 
\end{corollary} 
\begin{proof}
We have 
$\obsa$ positively determines $\PP$ iff 
$\obsa$ negatively determines $\BB\setminus\PP$ iff 
$\obsa$ negatively determines $\clop[\SF](\BB\setminus\PP)$ iff 
$\obsa$ positively determines $\BB \setminus \clop[\SF](\BB\setminus\PP)$. 
Then, the thesis follows by the following equalities
\[
\BB \setminus \clop[\SF](\BB\setminus\PP) = \inop[\coSF](\BB\setminus(\BB\setminus\PP)) = \inop[\coSF](\PP) 
\]
\end{proof}

\begin{proposition}
Let $\PP$ be a property on $\BB$ and $\obsa\in\Obs$ an observation. 
Then, $\amnt{\clop[\SF](\PP)}(\obsa) = \yv$ and $\amnt{\inop[\coSF](\PP)}(\obsa) = \nv$ 
if and only if, 
for every $\obsb\in\Obs$ such that $\obsa\obsord\obsb$, $\amnt\PP(\obsb) = \uv$. 
\end{proposition} 
\begin{proof}
Let us first prove the left-to-right implication. 
Since abstract monitors are impartial, we have that, for every $\obsb\in\Obs$ such that $\obsb\obsord\obsb$
$\amnt{\clop[\SF](\PP)}(\obsb) = \yv$ and $\amnt{\inop[\coSF](\PP)}(\obsb) = \nv$. 
Moreover, by \cref{thm:sf-nv,cor:cosf-yv}, 
we have that 
$\amnt{\clop[\SF](\PP)}(\obsb) = \yv$ implies $\amnt\PP(\obsb) \ne \nv$ and 
$\amnt{\inop[\coSF](\PP)}(\obsb) = \nv$ implies $\amnt\PP(\obsb) \ne \yv$. 
Therefore, we get $\amnt\PP(\obsb) = \uv$, as needed. 

Towards a proof of the other direction, 
observe that, for every $\obsb\in\Obs$ such that $\obsa\obsord\obsb$, since $\amnt\PP(\obsb) = \uv$, 
there are $\bha_1,\bha_2\in\BB(\obsb)$ such that 
$\bha_1\in\PP$ and $\bha_2\notin\PP$. 
Therefore, 
for every $\bha\in\BB(\obsa)$ and every observation $\obsb\bhord\bha$, 
since the observation structure is directed, there is $\obsc\bhord\bha$ such that $\obsa\obsord\obsc$ and $\obsb\obsord\obsc$, and, 
from what we observed above, 
there are $\bha_1,\bha_2\in\BB(\obsc)$ such that $\bha_1\in\PP$ and $\bha_2\notin\PP$, 
and, since $\obsa\obsord\obsc$, we have $\bha_1,\bha_2\in\BB(\obsb)$. 
By \cref{def:safety2} (and its dual), this implies that $\bha\in\clop[\SF](\PP)$ and $\bha\notin\inop\coSF](\PP)$, for every $\bha\in\BB(\obsa)$. 
Hence, we get $\amnt{\clop[\SF](\PP)}(\obsa) = \yv$ and $\amnt{\inop[\coSF](\PP)}(\obsa) = \nv$, as needed. 
\end{proof}

\begin{theorem}
Let $\PP$ be a property on $\BB$. 
Then, for every $\obsa\in\Obs$, there is $\obsb\in\Obs$  such that $\obsa\obsord\obsb$ and 
$\gamnt\PP(\obsb) \in \set{\yv,\nv,\nmv}$. 
\end{theorem}
\begin{proof}
Since $\clop[\SF](\PP)$ is monitorable (\cref{prop:monitorable-sf}), there is 
$\obsb\in\Obs$  such that $\obsa\obsord\obsb$ and $\amnt{\clop[\SF](\PP)}(\obsb) \ne \uv$. 
Moreover, since $\inop[\coSF](\PP)$ is monitorable (\cref{prop:co-safety,prop:monitorable-sf,prop:monitorable-compl}), there is 
$\obsc\in\Obs$  such that $\obsb\obsord\obsc$ and $\amnt{\inop[\coSF](\PP)}(\obsc) \ne \uv$. 
Then, since $\amnt{\clop[\SF](\PP)}$ is impartial, we have $\amnt{\clop[\SF](\PP)}(\obsc)\ne\uv$. 
Therefore, we get the thesis by definition of $\amnt\PP$. 
\end{proof}


\subsection{Linear Time $\nu$-Calculus}
\label{sec:app-ltnu-def}
\begin{definition}[Proof System]
	\label{def:ltnu_prfsys}
	The coinductive interpretation of the following inference system describes the proof system
	for LT$\nu$ terms using judgments of the form $\prf\tt$ for a term $\tt$.
	We use $\Ctx, \CtxD$ to range over sets of LT$\nu$ terms.
	\begin{mathpar}
		\inferrule[p-ax]{\mathstrut}
			{\prf{\pp_1,\dots,\pp_n,\co\pp_{n+1},\dots,\co\pp_{m},\Ctx}}
			~ \bigcup_{i=1}^n \psem{\pp_i} \cup 
			  \bigcup_{i=n+1}^m (\Itr \setminus \psem{\pp_i}) = \Itr
		\and
		\inferrule[p-top]{\mathstrut}{\prf{\top,\Ctx}}
		\and
		\inferrule[p-and]{\prf{\tt,\Ctx} \\ \prf{\ts,\Ctx}}{\prf{\tt \land \ts,\Ctx}}
		\and
		\inferrule[p-or]{\prf{\tt,\ts,\Ctx}}{\prf{\tt \lor \ts,\Ctx}}
		\and
		\inferrule[p-rec]{\prf{\tt\subst{\rec\var\tt}\var,\Ctx}}{\prf{\rec\var\tt,\Ctx}}
		\and
		\inferrule[p-next]{\prf\Ctx}{\prf{\Next\Ctx,\CtxD}}
	\end{mathpar}
	Where we write $\Next\Ctx$ for a set of LT$\nu$ terms of the form $\Next\tt_1,\dots,\Next\tt_n$.
	Note that the proof system is the same of that presented in\cite{DaxHL06} for cut-free derivations.
\end{definition}

\begin{definition}[LTS]
The following inductive inference system defines the labeled transition system for LT$\nu$ terms.
\begin{mathpar}
	\inferrule[r-top]{\mathstrut}{\top \red\ev \top}
	\and
	\inferrule[r-prop]{\mathstrut}{\pp \red\ev \top} ~ \ev \in \psem{\pp}
	\and
	\inferrule[r-coprop]{\mathstrut}{\co\pp \red\ev \top} ~ \ev \not\in \psem{\pp}
	\and
	\inferrule[r-next]{\mathstrut}{\Next\tt \red\ev \tt}
	\\
	\inferrule[r-and]{\tt \red\ev \tt' \\ \ts \red\ev \ts'}{\tt\land\ts \red\ev \tt'\land\ts'}
	\and
	\inferrule[r-or-l]{\tt \red\ev \tt'}{\tt\lor\ts \red\ev \tt'}
	\and
	\inferrule[r-or-r]{\ts \red\ev \ts'}{\tt\lor\ts \red\ev \ts'}
	\and
	\inferrule[r-rec]{\tt \subst{\rec X \tt}{X} \red\ev \tt'}{\rec X \tt \red\ev \tt'}
\end{mathpar}
Note that the LTS is inductively defined since we assumed that terms are \emph{contractive}.
\end{definition}

\begin{definition}
\label{def:ireds}
We define the predicate $\tt \ireds\itr$ meaning that the term $\tt$ reduces infinitely many times 
according to $\itr \in \Itr$.
We coinductively interpret the following rule
\[
\inferrule{
		\ts \ireds{\itr}
	}{
		\tt \ireds{\ev\itr}
	}~ \tt \move{e} \ts
\]
\end{definition}

\begin{app-thm}
	\label{thm:cmods_eq_ireds}
	Let $\itr \in \Itr$ and $\tt$ a LT$\nu$ term. Then $\itr \cmods \tt \iff \tt \ireds\itr$.
\end{app-thm}
The proof of \Cref{thm:cmods_eq_ireds} is in \Cref{sec:app-ltnu-corr}.
The proof of the iff part of \cref{thm:cmods_eq_ireds} is by coinduction. Additionally we have to prove that if
$\itr \cmods \tt$, then there exists $\ts$ such that $\tt \red{\itr_0} \ts$ 
and $\itr_{\geq 1} \cmods \ts$ (see \Cref{lem:prog+subj_red}).

\begin{example}
	Consider the LT$\nu$ term $\tt = \rec\var{(\pp \lor \co\pp) \land \Next\var}$.
	$\tt \neq \top$ but it is easy to see that $\csem\tt = \Itr$. 
	Hence, we provide a derivation inside the proof system.
	\[
	\begin{prooftree}
		\[
			\[
				\[
					\justifies
					\prf{\pp,\co\pp}
					\using\refrule{p-ax}
				\]
				\justifies
				\prf{\pp \lor \co\pp}
				\using\refrule{p-or}
			\]
			\[
				\[
					\vdots
					\justifies
					\prf\tt
					\using\refrule{p-rec}
				\]
				\justifies
				\prf{\Next\tt}
				\using\refrule{p-next}
			\]
			\justifies
			\prf{(\pp \lor \co\pp) \land \Next\tt}
			\using\refrule{p-and}
		\]
		\justifies
		\prf{\rec\var{(\pp \lor \co\pp) \land \Next\var}}
		\using\refrule{p-rec}
	\end{prooftree}
	\]
	Since the proof system is correct, 
	we deduce $\csem\tt = \Itr$.
	Thus, $\mnt_\tt(\ftr) = \yv$ for any $\ftr \in \Ftr$.
\end{example}


\subsection{Proof of \Cref{thm:csem_sf}}
\begin{definition}[Neg Semantics]
	\label{def:neg_csem}
	Let $\itr \in \Itr$. The negative semantics of a term is inductively defined by the following inference system.
\begin{mathpar}
	\inferrule[nc-bot]{\mathstrut}{\itr \not\cmods \bot}
	\and
	\inferrule[nc-prop]{\mathstrut}{\itr \not\cmods \pp} ~ \itr_0 \not\in \psem{\pp}
	\and
	\inferrule[nc-coprop]{\mathstrut}{\itr \not\cmods \co\pp} ~ \itr_0 \in \psem{\pp}
	\and
	\inferrule[nc-or]{\itr \not\cmods \tt \\ \itr \not\cmods \ts}{\itr \not\cmods \tt \lor \ts}
	\\
	\inferrule[nc-and-l]{\itr \not\cmods \tt}{\itr \not\cmods \tt \land \ts}
	\and
	\inferrule[nc-and-r]{\itr \not\cmods \ts}{\itr \not\cmods \tt \land \ts}
	\and
	\inferrule[nc-next]{\itr_{\geq 1} \not\cmods \tt}{\itr \not\cmods \Next\tt}
	\and
	\inferrule[nc-rec]{\itr \not\cmods \tt\subst {\rec X \tt} X}{\itr \not\cmods \rec X \tt}
\end{mathpar}
\end{definition}

\begin{app-thm}
	Let $\tt$ be a term of the calculus. Then $\csem\tt \in \SF$.
\end{app-thm}
\begin{proof}
By \Cref{def:safety} we have to prove that for each $\itr \in \Itr$ such that $\itr \not\cmods \tt$ there exists 
$\ftr \prefix \itr \in \Ftr$ such that for all $\itr' \in \Itr$ $\ftr\itr' \not\cmods \tt$.
We reason by induction on the judgment $\itr \not\cmods \tt$.

\proofrule{nc-bot}
	Then $\tt = \bot$.
	We conclude taking any prefix $\ftr$ and by applying \refrule{nc-bot}.

\proofrule{nc-prop}
	Then $\tt = \pp$ and $\itr_0 \not\in \psem\pp$. 
	We conclude taking $\ftr \eqdef \itr_0$ and applying \refrule{nc-prop}.

\proofrule{nc-coprop}
	Then $\tt = \co\pp$ and $\itr_0 \in \psem\pp$. 
	We conclude taking $\ftr \eqdef \itr_0$ and applying \refrule{nc-coprop}.

\proofrule{nc-or}
	Then $\tt = \tt' \lor \ts$, $\itr \not\cmods \tt'$ and $\itr \not\cmods\ts$.
	Using the induction hypotheses we deduce that there exist $\ftr_{\tt'} \prefix \itr$ and $\ftr_{\ts} \prefix \itr$ that
	negatively determine $\tt'$ and $\ts$ respectively.
	It must be the case that either $\ftr_{\tt'} \prefix \ftr_{\ts}$ or $\ftr_{\ts} \prefix \ftr_{\tt'}$.
	In the former case we take $\ftr \eqdef \ftr_{\ts}$ while in the second $\ftr \eqdef \ftr_{\tt'}$.
	We conclude applying \refrule{nc-or}.

\proofrule{nc-and-l}
	Then $\tt = \tt' \land \ts$ and $\itr \not\cmods \tt'$. Using the induction hypothesis we deduce that there exists
	$\ftr_{\tt'} \prefix \itr$ that negatively determines $\tt'$. 
	We conclude taking $\ftr \eqdef \ftr_{\tt'}$ and applying \refrule{nc-and-l}.

\proofrule{nc-and-r}
	Analogous to the previous case.

\proofrule{nc-next}
	Then $\tt = \Next\ts$ and $\itr_{\geq 1} \not\cmods \ts$.
	Using the induction hypothesis we deduce that there exists $\ftr_{\ts} \prefix \itr_{\geq 1}$ that negatively
	determines $\ts$. 
	We conclude taking $\ftr \eqdef \itr_0\ftr_{\ts}$ and applying \refrule{nc-next}.

\proofrule{nc-rec}
	Then $\tt = \rec X \ts$ and $\itr \not\cmods \ts \subst{\rec X \ts} X$.
	Using the induction hypothesis we deduce that there exists $\ftr_{\ts}$ that negatively determines $\ts \subst{\rec X \ts} X$.
	We conclude taking $\ftr \eqdef \ftr_{\ts}$ and applying \refrule{nc-rec}.
\end{proof}


\subsection{Proof of \Cref{thm:ireds-correct}}
\label{sec:app-ltnu-corr}
\begin{definition}[Rank of a term]
Let $\tt$ be a LT$\nu$ term. The function $\rank\tt$ computes the \emph{rank} of $\tt$.
\[
\begin{array}{lcllcl}
	\rank\top 			& = & 0	\qquad & \qquad	
	\rank\bot 			& = & 0 \\
	\rank\pp 			& = & 0	\qquad & \qquad	
	\rank{\co\pp} 		& = & 0 \\
	\rank{\tt\land\ts} 	& = & 1 +\max\set{\rank\tt,\rank\ts} \qquad & \qquad 
	\rank{\tt\lor\ts} 	& = & 1 +\max\set{\rank\tt,\rank\ts} \\
	\rank{\rec\var\tt} 	& = & 1 + \rank\tt \qquad & \qquad 
	\rank{\Next\tt} 	& = & 0
\end{array}
\]
\end{definition}

Intuitively, $\rank\tt$ measures the distance from $\tt$ to the first subformula with shape $\top, \bot, \pp, \co\pp$ or $\Next\ts$.
Note that the case $\rank\var$ is ruled out by the \emph{contractivity} assumption on terms.

\begin{lemma}
	\label{lem:rank_after_subst}
	Let $\tt$ be a LT$\nu$ term. Then $\rank\tt = \rank{\tt\subst\ts\var}$ for any $\ts$.
\end{lemma}
\begin{proof}
By induction on $\tt$ and by cases on its shape.

\proofcase{Case $\tt = \bot$, $\tt = \top$, $\tt = \pp$, $\tt = \co\pp$}
We conclude $\rank\tt = \rank{\tt\subst\ts\var} = 0$ since $\var \not\in \fn\tt$.

\proofcase{Case $\tt = \tt_1 \land \tt_2$}
Using the induction hypotheses we deduce 
\begin{itemize}
\item $\rank{\tt_1} = \rank{{\tt_1}\subst\ts\var}$
\item $\rank{\tt_2} = \rank{{\tt_2}\subst\ts\var}$
\end{itemize}
Hence we deduce
\[
\max\set{\rank{\tt_1},\rank{\tt_2}} = \max\set{\rank{{\tt_1}\subst\ts\var},\rank{{\tt_2}\subst\ts\var}}
\]
We conclude $\rank\tt = \rank{\tt\subst\ts\var}$.

\proofcase{Case $\tt = \tt_1 \lor \tt_2$}
Analogous to the previous case.

\proofcase{Case $\tt = \rec\vary{\tt'}$ and $\var \neq \vary$}
Using the induction hypothesis we deduce $\rank{\tt'} = \rank{{\tt'}\subst\ts\var}$.
We conclude $\rank\tt = \rank{\tt\subst\ts\var}$.

\proofcase{Case $\tt = \Next\ts$}
We conclude $\rank\tt = \rank{\tt\subst\ts\var} = 0$ by definition of $\rank\tt$.
\end{proof}

\begin{lemma}
	\label{lem:prog+subj_red}
	Let $\ev \in \Events, \itr \in \Itr$ and $\tt$ a LT$\nu$ term. 
	Then $\ev\itr \cmods \tt$ implies $\tt \red\ev \ts$ for some $\ts$ and $\itr \cmods \ts$.
\end{lemma}
\begin{proof}
By complete arithmetic induction on $\rank\tt$ and by cases on $\ev\itr \cmods \tt$.

\proofcase{Case $\rank\tt = 0$}
Then it must be the case that $\ev\itr \cmods \tt$ is the consequence of 
\refrule{c-top} or \refrule{c-prop} or \refrule{c-coprop} or \refrule{c-next}.

	\proofcase{$\qquad$Case \refrule{c-top}, \refrule{c-prop}, \refrule{c-coprop}}
	Then we take $\ts \eqdef \top$ noting that $\tt \red\ev \top$ by either 
	\refrule{r-top} or \refrule{r-prop} or \refrule{r-coprop}.
	We conclude with one application of \refrule{c-top}.
	
	\proofcase{$\qquad$Case \refrule{c-next}}
	Then $\tt = \Next\ts \red\ev \ts$ by \refrule{r-next} and $\itr \cmods \ts$.
	There is nothing left to prove.
	
\proofcase{Case $\rank\tt > 0$}
Then it must be the case that $\ev\itr \cmods \tt$ is the consequence of 
\refrule{c-and} or \refrule{c-or-l} or \refrule{c-or-r} or \refrule{c-rec}.

	\proofcase{$\qquad$Case \refrule{c-and}}
	Then $\tt = \tt_1 \land \tt_2, \ev\itr \cmods \tt_1$ and $\ev\itr \cmods \tt_2$.
	Note that $\rank{\tt_1} < \rank\tt$ and $\rank{\tt_2} < \rank\tt$.
	Using the induction hypotheses we deduce that there exist $\ts_1, \ts_2$ such that
	\begin{itemize}
	\item $\tt_1 \red\ev \ts_1$ and $\itr \cmods \ts_1$
	\item $\tt_2 \red\ev \ts_2$ and $\itr \cmods \ts_2$
	\end{itemize}		
	We conclude taking $\ts \eqdef \ts_1 \land \ts_2$ and applying \refrule{r-and} and \refrule{c-and}.

	\proofcase{$\qquad$Case \refrule{c-or-l}}
	Then $\tt = \tt_1 \lor \tt_2$ and $\ev\itr \cmods \tt_1$.
	Note that $\rank{\tt_1} < \rank\tt$.
	Using the induction hypothesis we deduce that there exists $\ts_1$ such that
	$\tt_1 \red\ev \ts_1$ and $\itr \cmods \ts_1$.
	We conclude taking $\ts \eqdef \ts_1$ and applying \refrule{r-or-l}.

	\proofcase{$\qquad$Case \refrule{c-or-r}}
	Analogous to the previous case.
	
	\proofcase{$\qquad$Case \refrule{c-rec}}
	Then $\tt = \rec\var\tt'$ and $\ev\itr \cmods \tt'\subst{\rec\var\tt'}\var$.
	Note that $\rank{\tt'} < \rank\tt$.
	From \cref{lem:rank_after_subst} we deduce $\rank{\tt'} = \rank{\tt'\subst{\rec\var\tt'}\var} < \rank\tt$.
	Using the induction hypothesis we deduce that there exists $\ts'$ such that 
	$\tt'\subst{\rec\var\tt'}\var \red\ev \ts'$ and $\itr \cmods \ts'$.
	We conclude taking $\ts \eqdef \ts'$ and applying \refrule{r-rec}.
\end{proof}

\begin{app-thm}
	Let $\itr \in \Itr$ and $\tt$ a LT$\nu$ term. Then $\itr \cmods \tt$ implies $\tt \ireds\itr$.
\end{app-thm}
\begin{proof}
The proof is by coinduction. Let $\spec = \set{(\ev\itr, \tt) \mid \ev\itr \cmods \tt}$, we prove that $\spec$ is
\emph{consistent} with respect to $\ireds{}$.
In particular, if $\ev\itr \cmods \tt$ we have to prove that there exists $\ts$ 
such that $\tt \red\ev \ts$ and $\itr \cmods \ts$.
We conclude applying \cref{lem:prog+subj_red}.
\end{proof}

\begin{lemma}
	\label{lem:and_reduction}
	Let $\itr \in \Itr$ and $\tt,\ts$ LT$\nu$ terms.
	Then $\tt \land \ts \ireds\itr$ implies, $\tt \ireds\itr$ and $\ts \ireds\itr$.
\end{lemma}
\begin{proof}
We prove that $\tt \land \ts \ireds\itr$ implies $\tt \ireds\itr$. The other proof is analogous.
The proof is by coinduction. Let $\spec = \set{(\tt, \ev\itr) \mid \exists\ts.\tt \land \ts \ireds{\ev\itr}}$. 
We prove that $\spec$ is \emph{consistent} with respect to $\ireds{}$.

From \refrule{r-and} we deduce that there exist $\tt', \ts'$ such that
\begin{itemize}
\item $\tt \land \ts \red\ev \tt' \land \ts'$  
\item $\tt \red\ev \tt'$ and $\ts \red\ev \ts'$
\item $\tt' \land \ts' \ireds\itr$
\end{itemize}
We have to prove $(\tt', \itr) \in \spec$.
We conclude by observing that there exists $\ts'$ such that $\tt' \land \ts' \ireds\itr$.
\end{proof}

\begin{app-thm}
	Let $\itr \in \Itr$ and $\tt$ a LT$\nu$ term. Then $\tt \ireds\itr$ implies $\itr \cmods \tt$.
\end{app-thm}
\begin{proof}
The proof is by coinduction. Let $\spec = \set{(\ev\itr, \tt) \mid \tt \ireds{\ev\itr}}$, we prove that $\spec$ is
\emph{consistent} with respect to $\cmods$.
We proceed by cases on $\tt$.

\proofcase{Case $\tt = \top$}
	Then $\ev\itr \cmods \top$ by \refrule{c-top}.

\proofcase{Case $\tt = \pp$}
	From \refrule{r-prop} we deduce $\pp \red\ev \top$ and $\ev \in \psem\pp$. 
	Then $\ev\itr \cmods \pp$ is the consequence of \refrule{c-prop}.

\proofcase{Case $\tt = \co\pp$}
	From \refrule{r-coprop} we deduce $\co\pp \red\ev \top$ and $\ev \not\in \psem\pp$. 
	Then $\ev\itr \cmods \co\pp$ is the consequence of \refrule{c-coprop}.

\proofcase{Case $\tt = \tt_1 \land \tt_2$}
	From \refrule{r-and} we deduce that there exist $\ts_1, \ts_2$ such that
	\begin{itemize}
	\item $\tt_1 \land \tt_2 \red\ev \ts_1 \land \ts_2$
	\item $\tt_1 \red\ev \ts_1$ and $\tt_2 \red\ev \ts_2$
	\item $\ts_1 \land \ts_2 \ireds\itr$
	\end{itemize}
	Then $\ev\itr \cmods \tt_1 \land \tt_2$ is the consequence of \refrule{c-and}.
	We have to prove $\tt_1 \ireds{\ev\itr}$ and $\tt_1 \ireds{\ev\itr}$.
	Using \cref{lem:and_reduction} we deduce $\ts_1 \ireds\itr$ and $\ts_2 \ireds\itr$.
	We conclude observing that $\tt_1 \red\ev \ts_1$ and $\tt_2 \red\ev \ts_2$.
	
\proofcase{Case $\tt = \tt_1 \lor \tt_2$}
	Note that $\tt_1 \lor \tt_2$ can reduce with either \refrule{r-or-l} of \refrule{r-or-r}.
	We consider the case in which \refrule{r-or-l} is used, the other one is symmetric.
	From \refrule{r-or-l} we deduce that there exists $\tt_1'$ such that
	\begin{itemize}
	\item $\tt_1 \lor \tt_2 \red\ev \tt_1'$
	\item $\tt_1 \red\ev \tt_1'$
	\item $\tt_1' \ireds\itr$
	\end{itemize}
	Then it must be the case that $\ev\itr \cmods \tt_1 \lor \tt_2$ is the consequence of\refrule{c-or-l}.
	We have to prove that $\tt_1 \ireds{\ev\itr}$.
	The proof is straightforward since $\tt_1 \red\ev \tt_1'$ and $\tt_1' \ireds\itr$.
	
\proofcase{Case $\tt = \rec\var\ts$}
	From \refrule{c-rec} we deduce that
	\begin{itemize}
	\item $\rec\var\ts \red\ev \tt'$
	\item $\ts\subst{\rec\var\ts}\var \red\ev \tt'$
	\item $\tt' \ireds\itr$
	\end{itemize}		
	Then $\ev\itr \cmods \rec\var\ts$ is the consequence of \refrule{c-rec}.
	We have to prove $\ts\subst{\rec\var\ts}\var \ireds{\ev\itr}$
	The proof is straightforward since $\ts\subst{\rec\var\ts}\var \red\ev \tt'$ and $\tt' \ireds\itr$.

\proofcase{Case $\tt = \Next\ts$}
	From \refrule{r-next} we deduce $\Next\ts \red\ev \ts$ and $\ts \ireds\itr$.
	Then $\ev\itr \cmods \Next\ts$ is the conclusion of \refrule{c-next}.
	We have to prove that $\ts \ireds\itr$.
	There is nothing left to prove.
\end{proof}


\subsection{Proof of \Cref{thm:ltl-ltnu}}
Since it is always possible to provide a \Buchi automaton starting from a linear temporal logic formula such that the accepted language coincides with the semantics of the formula, we propose an algorithm that acts directly on \Buchi automata.
First we recall the definition of a \Buchi automaton.

\begin{definition}[\Buchi Automaton]

A \Buchi Automaton $\Automaton$ is a tuple 
$\tuple{\States, \Alphabet , \trans , \IStates , \FStates}$ such that
\begin{itemize}
\item $\States$ is a finite set of states
\item $\Alphabet$ is a finite set called alphabet
\item $\trans: \States \times \Alphabet \rightarrow \wp(\States)$ is a transition relation
\item $\IStates \subseteq \States$ is the set of initial states
\item $\FStates \subseteq \States$ is the set of final states
\end{itemize}
We use $\st$ to denote an element of $\States$ and we denote $\Language\Automaton$ the language accepted by the automaton $\Automaton$.
We fix $\Alphabet = \PFstar{\AP}$ and we use $\act,\actb$ to denote an element of $\Alphabet$.
We write $\st \move\act \st'$ for $\st' \in \trans(\st,\act)$.
We write $\st \moves{\ftr} \st'$ when $\st$ reduces in finite steps to $\st'$ according to $\ftr$.
We write $\st \imoves{\itr}$ when $\st$ reduces infinitely many times according to $\itr$
(coinductively defined as for \Cref{def:ireds}).
$\ftr$ and $\itr$ denote respectively finite and infinite sequence of elements of $\Alphabet$.
$\itr \in \Language\Automaton$ if $\st \imoves\itr$ for some $\st \in \IStates$ and such that the reduction passes
through some final state in $\FStates$ infinitely often.
\end{definition}

\begin{property}[Assumptions]
\label{prop:nba_assumptions}
	We assume that input NBAs satisfy the following properties:
	\begin{itemize}
	\item For each $\st \in \FStates$, $\st$ lies in a cycle
	\item For all $\st \in \States$, $\st \moves\itr \st'$ implies $\st' \moves{\itr'} \st_f$ 
	for some $\st_f \in \FStates, \itr, \itr'$
	\end{itemize}
\end{property}

\begin{lemma}[Emptyness]
	\label{lem:empty_automaton}
	Let $\Automaton = \tuple{\States, \Alphabet , \trans , \IStates , \FStates}$.
	According to the assumption we make, we obtain $\Language\Automaton = \emptyset$ iff $\States = \emptyset$.
\end{lemma}
\begin{proof}
	The \emph{if} part is obvious.
	To prove that if $\Language\Automaton = \emptyset$ then $\States = \emptyset$ it is sufficient to observe
	that in such an automaton either the paths (from an initial state) do not reach some final state or 
	some final state in reachable but it does not belong to a cycle.
	Such kind of NBAs are ruled out by \cref{prop:nba_assumptions}.  
\end{proof}

The algorithm to encode a \Buchi automaton into a LT$\nu$ term is described in \Cref{alg}. Next, we show that such encoding 
produces the safety approximation of the property described by the input automaton.

\begin{algorithm}
\caption{Encoding}
\label{alg}
\begin{algorithmic}[1]
\item[]
\State Let $\Automaton = \tuple{\States, \Alphabet , \trans , \IStates , \FStates}$
\State Assume a variable $\var_\st$ for each $\st \in \States$
\item[]
\Procedure{$T$}{$\st,\SSet$}
	\If{$\st \in \SSet$} $X_q$
	\Else
		\State $\rec
			{\var_\st}
			{\lor \set{
				$\Call{$T$}{$\act$}$ \land \Next$\Call{$T$}{$\st',\SSet\cup\set{\st}$}$ 
				\mid 
				\act \in \Alphabet, \st' \in \trans(\st,\act)}}$
	\EndIf
\EndProcedure
\item[]
\State \Call{$T$}{$\act$} = $\land \set{\pp \mid \pp \in \act} \land \set{\co\pp \mid \pp \not\in \act}$
\State \Call{$T$}{$\Automaton$} = $\lor \set{$\Call{$T$}{$\st,\emptyset$}$\mid \st \in \IStates}$
\end{algorithmic}
\end{algorithm}

Let $\Automaton$ be a \Buchi automaton satisfying \cref{prop:nba_assumptions}.
We want to prove that
\begin{itemize}
\item $\clop[\SF](\Language\Automaton) \subseteq \csem{\Talg\Automaton}$ (\Cref{thm:comp_lang_in_csem})
\item $\csem{\Talg\Automaton} \subseteq \clop[\SF](\Language\Automaton)$ (\Cref{thm:csem_in_comp_lang})
\end{itemize}


First, we define the predicate $\drv\_\_\_$.

\begin{definition}
	Let $\Automaton = \tuple{\States, \Alphabet , \trans , \_ , \_}$.
	We consider the inductive interpretation of the following inference system that derives
	judgments of the form $\drv\SSet\st\tt$ where $\SSet$ is a set of states, $\st \in \States$ and
	$\tt$ a LT$\nu$ term.
	\begin{mathpar}
		\inferrule[drv-var]{\mathstrut}{\drv\SSet\st{\var_{\st}}} ~ \st \in \SSet
		\and
		\inferrule[drv-rec]
			{
			\forall \act,\st'.\act \in \Alphabet,\st'\in\trans(\st,\act). 
				\drv{\SSet\cup\set\st}{\st'}{\tt_{\st'}^\act}
			}{
			\drv{\SSet}{\st}
				{
				\rec{\var_\st}{\lor\set{\Tact\act \land \Next\tt_{\st'}^\act
					\mid \act,\st'.\act \in \Alphabet,\st'\in\trans(\st,\act)}}
				}
			}
	\end{mathpar}
\end{definition}

\begin{lemma}
	\label{lem:talg_drv}
	Let $\Automaton$ be a \Buchi automaton, $\st$ one of its states and $\SSet$ a set of states.
	Then $\drv\SSet\st{\Tqalg\st\SSet}$.
\end{lemma}
\begin{proof}
	By induction over $\Tqalg\_\_$ and applying \refrule{drv-var} as soon as possible.
\end{proof}

\begin{lemma}[Weakening]
	\label{lem:drv_weaken}
	Let $\st$ be a state of an automaton, $\tt$ a LT$\nu$ term and $\SSet,\SSet'$ two sets of states.
	If $\drv\SSet\st\tt$ then $\drv{\SSet\cup\SSet'}\st\tt$.
\end{lemma}
\begin{proof}
	By induction over $\drv\SSet\st\tt$.
\end{proof}

\begin{lemma}
	\label{lem:drv_subst}
	Let $\st,\st'$ be states of an automaton, $\tt,\tt'$ LT$\nu$ terms and $\SSet$ a set of states.
	If $\drv{\SSet\cup\set\st}{\st'}{\tt'}$ and $\drv\SSet\st\tt$, 
	then $\drv\SSet{\st'}{\tt'\subst\tt{\var_q}}$.
\end{lemma}
\begin{proof} 
	By induction on $\drv{\SSet\cup\set\st}{\st'}{\tt'}$ and using \cref{lem:drv_weaken}.
	
	\proofcase{Case \refrule{drv-var} and $\st \in \SSet$}
	Then $\tt' = \var_{\st'}$ and $\drv\SSet{\st'}{\tt'\subst\tt{\var_q}}$
	since $\var_\st \not\in \fn{\var_{\st'}}$.
	
	\proofcase{Case \refrule{drv-var} and $\st = \st'$}
	Then $\tt' = \var_\st$ and $\tt'\subst\tt{\var_q} = \tt$.
	There is nothing left to prove.
	
	\proofcase{Case \refrule{drv-rec}}
	Then
	\begin{itemize}
	\item $\tt' = \rec{\var_{\st'}}{\lor\set{\Tact\act \land \Next\tt_{\st''}^\act
					\mid \act,\st''.\act \in \Alphabet,\st''\in\trans(\st',\act)}}$
	\item $\forall \act,\st''.\act \in \Alphabet,\st''\in\trans(\st',\act). 
				\drv{\SSet\cup\set{\st,\st'}}{\st''}{\tt_{\st''}^\act}$
	\end{itemize}
	Using \cref{lem:drv_weaken} we deduce $\drv{\SSet\cup\set{\st'}}\st\tt$.
	Using the induction hypotheses we deduce 
	$\forall \act,\st''.\act \in \Alphabet,\st''\in\trans(\st',\act). 
				\drv{\SSet\cup\set{\st'}}{\st''}{\tt_{\st''}^\act \subst\tt{\var_q}}$.
	We conclude applying \refrule{drv-rec}.
\end{proof}

\begin{lemma}
	\label{lem:buchi_red}
	Let	$\Automaton$ be a \Buchi automaton, $\st,\st'$ two of its states and $\tt$ LT$\nu$ term.
	If $\st \move\act \st'$ for some $\act \in \Alphabet$ and $\drv\emptyset\st\tt$
	then $\tt \red\act \top \land \top \land \dots \land \top \land \tt'$ for some $\tt'$
	such that $\drv\emptyset{\st'}{\tt'}$.
\end{lemma}
\begin{proof}
	First, we prove that, if $\st \move\act \st'$ for some $\act$, then there exists $\tt'$ such that
	$\tt \red\act \top \land \top \land \dots \land \tt'$ such that
	$\drv{\set\st}{\st'}{\tt'}$.
	
	It must be the case that $\drv\emptyset\st\tt$ is the consequence of \refrule{drv-rec}.
	Hence, we deduce 
	\begin{itemize}
	\item $\tt = \rec{\var_q}{\lor\set{\Tact\actb \land \Next\tt_{\st'}^\actb
	\mid \actb,\st'.\actb \in \Alphabet,\st'\in\trans(\st,\actb)}}$
	\item $\forall \actb,\st'.\actb \in \Alphabet,\st'\in\trans(\st,\actb).\drv{\set\st}{\st'}{\tt_{\st'}^\actb}$
	\end{itemize}
	
	It is easy to see that 
	$\tt \red\act \top \land \top \land \dots \land \tt_{\st'}^{\act} \subst\tt{\var_q}$
	by applying \refrule{r-rec}, \refrule{r-or-l}, \refrule{r-or-r}, \refrule{r-next},
	\refrule{r-prop} and \refrule{r-coprop}.
	
	Now we have to prove that $\drv\emptyset{\st'}{\tt_{\st'}^{\act} \subst\tt{\var_q}}$.

	We conclude using \cref{lem:drv_subst} and $\drv{\set\st}{\st'}{\tt_{\st'}^\act}$.
\end{proof}

\begin{lemma}
	\label{lem:stlang_in_csem}
	Let $\Automaton$ be a \Buchi automaton satisfying \cref{prop:nba_assumptions} 
	and $\st$ one of its states and $\tt$ a LT$\nu$ term. 
	If $\drv\emptyset\st\tt$, then $\Language\st \subseteq \csem\tt$. 
\end{lemma}
\begin{proof}
 	We have to prove that if $\drv\emptyset\st\tt$, then  for all $\itr \in \Itr$, 
 	$\itr \in \Language\st$ implies $\itr \cmods \tt$.
 	By \cref{thm:cmods_eq_ireds} this is equivalent to prove that $\tt \ireds\itr$.
 	The proof is by coinduction.
 	
 	Let $\spec = \set{(\act\itr, \tt) \mid \drv\emptyset\st\tt,\act\itr \in \Language\st}$. 
 	We have to prove that $\spec$ is \emph{consistent} with respect to $\cmods$.
 	In particular, we have to prove that $\tt \move\act \ts$ 
 	for some $\ts$ and $(\itr,\ts) \in \spec$.
 	
 	From $\act\itr \in \Language\st$ we deduce
 	\begin{itemize}
 	\item $\st \move\act \st' \imoves{\itr}$ for some $\st'$
 	\item $\itr \in \Language{\st'}$
	\end{itemize}
 	Using \cref{lem:buchi_red} we deduce 
 	$\tt \move\act \top \land \top \land \dots \land \ts$ and $\drv\emptyset{\st'}\ts$.
 	We conclude observing that 
 	$\csem{\top \land \top \land \dots \land T_q(\st', \emptyset)} = \csem{T_q(\st', \emptyset)}$.
\end{proof}

\begin{lemma}
	\label{lem:lang_in_csem}
	Let Let $\Automaton = \tuple{\_, \_ , \_ , \IStates , \_}$ 
	be a \Buchi automaton satisfying \cref{prop:nba_assumptions}. 
	Then $\Language\Automaton \subseteq \csem{\Talg\Automaton}$. 
\end{lemma}
\begin{proof}
	By definition, $\Language\Automaton = \bigcup\set{\Language\st \mid \st \in \IStates}$
	and $\csem{\Talg\Automaton} = \bigcup\set{\csem{\Tqalg\st\emptyset} \mid \st \in \IStates}$.
	It is sufficient to prove that $\Language\st \subseteq \Tqalg\st\emptyset$ for all $\st \in \IStates$.
	We conclude by applying \cref{lem:talg_drv} and \cref{lem:stlang_in_csem}.
\end{proof}

\begin{app-thm}
	\label{thm:comp_lang_in_csem}
	Let $\Automaton$ be a \Buchi automaton satisfying \cref{prop:nba_assumptions}.
	
	Then $\clop[\SF](\Language\Automaton) \subseteq \csem{\Talg\Automaton}$.
\end{app-thm}
\begin{proof}
	By definition, $\clop[\SF](\Language\Automaton)$ is the intersection of all the 
	safety properties that include $\Language\Automaton$.
	We conclude using \cref{lem:lang_in_csem} and \cref{thm:csem_sf}.
\end{proof}

\begin{lemma}
	\label{lem:drv_st_red}	
	Let $\st$ be a state of a \Buchi automaton, $\itr \in \Itr$ and $\tt$ a LT$\nu$ term.
	If $\drv\emptyset\st\tt$ and $\itr \cmods \tt$, then $\st \ireds\itr$. 
\end{lemma}
\begin{proof}
	By coinduction considering $\spec = \set{(\act\itr,\st) \mid \drv\emptyset\st\tt, \act\itr \cmods \tt}$.
	We have to prove that $\st \move\act \st'$ for some $\st'$ and $(\itr, \st') \in \spec$.
	
	First, it must be the case that $\drv\emptyset\st\tt$ is the consequence of \refrule{drv-rec}.
	Hence, we deduce
	\begin{itemize}
	\item $\tt = \rec{\var_q}{\lor\set{\Tact\actb \land \Next\tt_{\st'}^\actb
					\mid \actb,\st'.\actb \in \Alphabet,\st'\in\trans(\st,\actb)}}$
	\item $\forall \actb,\st'.\actb \in \Alphabet,\st'\in\trans(\st,\actb). 
				\drv{\set\st}{\st'}{\tt_{\st'}^\actb}$
	\end{itemize}
	Hence $\st \red\actb \st_{\st'}$ for some $\actb$.
	
	Then from $\act\itr \cmods \tt$ we deduce 
	$\act\itr \mods \Tact\actb \land \Next \tt_{\st'}^\actb \subst\tt{\var_\st}$
	since \refrule{c-rec}, \refrule{c-or-l}, \refrule{c-or-r} have been applied to derive $\act\itr \cmods \tt$. 
	From \refrule{c-and} and \refrule{c-next} we deduce 
	\begin{itemize}
	\item $\act\itr \cmods \Tact\actb$
	\item $\act\itr \cmods \Next \tt_{\st'}^\actb \subst\tt{\var_\st}$ and $\itr \cmods \tt_{\st'}^\actb \subst\tt{\var_\st}$
	\end{itemize}
	By definition, $\Tact\actb = \land \set{\pp \in \actb} \land \set{\co\pp \mid \pp \not\in \actb}$.
	From \refrule{c-and}, \refrule{c-prop} and \refrule{c-coprop} we deduce
	\begin{itemize}
	\item $\act\itr \cmods \pp$ for all $\pp \in \actb$
	\item $\act\itr \cmods \co\pp$ for all $\pp \not\in \actb$
	\end{itemize}
	Hence we deduce $\act = \actb$.
	
	Finally, from $\drv{\set\st}{\st'}{\tt_{\st'}^\actb}$ we deduce $\drv\emptyset{\st'}{\tt_{\st'}^\actb \subst\tt{\var_\st}}$.
	
	We conclude $(\itr,\st') \in \spec$.
\end{proof}

\begin{app-thm}
	\label{thm:csem_in_comp_lang}
	Let $\Automaton$ be a \Buchi automaton satisfying \cref{prop:nba_assumptions}.
	
	Then $\csem{\Talg\Automaton} \subseteq \clop[\SF](\Language\Automaton)$.
\end{app-thm}
\begin{proof}
	If $\itr \cmods \Talg\Automaton$, then by definition of $\Talg\_$ it must be the case that $\itr \cmods \Tqalg\st\emptyset$
	for some $\st$ initial state of $\Automaton$.
	
	From \cref{lem:talg_drv} we deduce $\drv\emptyset\st{\Tqalg\st\emptyset}$.
	
	From \cref{lem:drv_st_red} on $\drv\emptyset\st{\Tqalg\st\emptyset}$ and $\itr \cmods \Tqalg\st\emptyset$
	we deduce $\st \imoves\itr$.
	
	Hence, by \cref{cor:sf-alt} we have to prove that for each $\ftr \prefix \itr \in \Ftr$ there exists $\itr' \in \Itr$
	such that $\ftr\itr' \in \Language\Automaton$.
	This is always possible since $\Automaton$ satisfies \cref{prop:nba_assumptions}.
\end{proof}


\subsection{Linear Temporal Logic}
\label{sec:app-ltl}
The semantics of LTL formulae is specified by a satisfation relation $\itr,i\lmods\ff$ (see \Cref{def:ltl_semantics}), 
stating that  an infinite trace $\itr$ satisfies the formula $\ff$ starting from the index $i\in\N$. We write $\itr \lmods \ff$ iff $\itr,0 \lmods \ff$.
The semantics of a formula $\ff$ is then defined as $\lsem\ff = \set{\itr\in\Itr\mid \itr\lmods\ff}$. 

\begin{definition}[LTL Syntax]
\label{def:ltl_syntax}
The set of \emph{negation free} LTL formulae is inductively defined by the grammar
\[
	\ff,\fp \Coloneqq 	\top \mid \bot \mid \pp \mid \co\pp \mid \ff\land\fp \mid \ff\lor\fp 
						\mid \ff\until\fp \mid \ff\release\fp \mid \Next\ff 
\]
We use the following abbreviations
\[
\begin{array}{lcl}
	\F\ff 				& \eqdef & 	\top\until\ff
	\\
	\G\ff 				& \eqdef & 	\neg\F\neg\ff
\end{array}
\]
\end{definition}

\begin{definition}[LTL Semantics]
\label{def:ltl_semantics}
\[
\begin{array}{llcl}
	\defrule{l-top} 	& \itr,i \lmods \top \\
	\defrule{l-prop} 	& \itr,i \lmods \pp 			& \iff & \itr_i \in \psem{\pp} \\
	\defrule{l-coprop} 	& \itr,i \lmods \co\pp 			& \iff & \itr_i \not\in \psem{\pp} \\
	\defrule{l-and} 	& \itr,i \lmods \ff\land\fp 	& \iff & \itr,i \lmods \ff$ \bf{and} $\itr,i \lmods \fp \\
	\defrule{l-or} 		& \itr,i \lmods \ff\lor\fp 		& \iff & \itr,i \lmods \ff$ \bf{or} $\itr,i \lmods \fp \\
	\defrule{l-until} 	& \itr,i \lmods \ff\until\fp 	& \iff & \exists k \geq i. \itr,k \lmods \fp$ \bf{and} $\forall i \le l < k. \itr,l \lmods \ff \\
	\defrule{l-release} & \itr,i \lmods \ff\release\fp 	& \iff & \forall k \geq i. \itr,k \lmods \fp$ \bf{or} $\\
						&								&	   & \exists k \geq i. \itr,k \lmods \ff$ \bf{and} $\forall i \le l \le k. \itr,l \lmods \fp \\
	\defrule{l-next} 	& \itr,i \lmods \Next\ff 		& \iff & \itr,i+1 \lmods \ff
\end{array}
\]
\end{definition}

\end{document}